\documentclass[usletter,12pt]{article}
\pdfoutput=1 

\usepackage{jheppub} 

\usepackage[T1]{fontenc} 

\usepackage{amsmath,amsfonts,epsf}
\usepackage{amssymb}
\usepackage{graphicx}
\usepackage{grffile}
\usepackage{dsfont}
\usepackage{hyperref}
\input epsf

\def\bC {\mathds{C}}

\newcommand{\e}{\epsilon}
\newcommand{\m}{\mu}

\newcommand{\s}{\sigma}
\newcommand{\ld}{\lambda}

\newcommand{\A}{{\alpha}}
\newcommand{\B}{{\beta}}
\newcommand{\C}{{\gamma}}
\newcommand{\D}{{\delta}}
\newcommand{\da}{{\dot\alpha}}
\newcommand{\db}{{\dot\beta}}

\newcommand{\pd}{\partial}

\newcommand{\vev}[1]{{\left< {#1} \right>}}
\newcommand{\bra}[1]{{\left< {#1} \right|}}
\newcommand{\ket}[1]{{\left| {#1} \right>}}

\numberwithin{equation}{section}

\begin{document}

\title{Constraining conformal field theories with a higher spin symmetry in $d=4$}
\author[]{Vasyl Alba,}
\author[]{Kenan Diab}

\affiliation[]{Department of Physics 
\\ 
Jadwin Hall, Princeton University,\\
Princeton, NJ 08544 USA}

\emailAdd{valba@princeton.edu}
\emailAdd{kdiab@princeton.edu}

\abstract{We study unitary conformal field theories with a unique stress tensor and at least one
higher-spin conserved current in four dimensions.  We prove that every such theory contains an
infinite number of higher-spin conserved currents of arbitrarily high spin, and that Ward identities
generated by the conserved charges of these currents suffice to completely fix the correlators of
the stress tensor and the conserved currents to be equal to one of three free field theories: the
free boson, the free fermion, and the free vector field.  This is a generalization of the result
proved in three dimensions by Maldacena and Zhiboedov \cite{Maldacena:2011jn}.}

\arxivnumber{1307.8092}
\maketitle

\section*{Introduction}\label{introduction}
\addcontentsline{toc}{section}{Introduction}
Characterizing the theories dual to Vasiliev's higher-spin gauge theories in anti
de-Sitter space\cite{Konstein:2000bi}\cite{Vasiliev:2003ev}\cite{Vasiliev:2004qz} under the AdS/CFT
correspondence\cite{Maldacena:1997re}\cite{Gubser:1998bc}\cite{Witten:1998qj} has been a topic of
active research for over ten years, starting from the conjecture of Klebanov and Polyakov that
Vasiliev's theory in four dimensions is dual to the critical $O(N)$ vector model in three 
dimensions\cite{Klebanov:2002ja}\cite{Sezgin:2003pt}.  Under general principles of AdS/CFT, we
expect that the conformal field theory duals to Vasiliev's theories (when given appropriate boundary
conditions) should also have higher-spin symmetry, so it is natural to try to classify all
higher-spin conformal field theories.  In the case of CFT's in three dimensions, this task has
already been accomplished by Maldacena and Zhiboedov\cite{Maldacena:2011jn}, who showed that unitary
conformal field theories with a unique stress tensor and a higher-spin current are essentially free
in three dimensions.  This can be viewed as an analogue of the Coleman-Mandula
theorem\cite{Coleman:1967ad}\cite{Haag:1974qh}, which states that the maximum spacetime symmetry of
theories with a nontrivial S-matrix is the super-Poincare group.  

In this paper, we will prove a four-dimensional analogue of the Coleman-Mandula theorem for generic
conformal field theories.  We will show that in any unitary conformal field theory with a symmetric
conserved current of spin larger than $2$ and a unique stress tensor in four dimensions, all
correlation functions of symmetric currents of the theory are equal to the correlation functions of
a free field theory - either the free boson, the free fermion, or the free vector field. However, a
recent paper by Boulanger, Ponomarev, Skvortsov, and Taronna \cite{Boulanger:2013zza} strongly
indicates that all the algebras of higher-spin charges that are consistent with conformal symmetry
are not only Lie  algebras but associative. Hence, they are all reproduced by the universal
enveloping construction of \cite{Boulanger:2011se} with the conclusion that  any such algebra must
contain a symmetric higher-spin current. This implies that our result should be true even after relaxing our assumption that
the higher-spin current is symmetric. The argument is structured as follows:

In the first two sections, we will develop two technical tools which help us solve
certain Ward identities:
\begin{description}
\item[In section \ref{lightcone-limits},]we will define a particular limit of three-point functions
of symmetric conserved currents called \textit{lightcone limits}.  We will show that such correlation
functions behave essentially like correlation functions of a free theory in these limits, enabling
us to translate complicated Ward identities of the full theory into simpler ones involving only free
field correlators.  
\item[In section \ref{form-factors},]we will explain how one can use the spinor-helicity formalism
to convert Fourier-space matrix elements of conserved currents into simple polynomials. This will
allow us to further simplify Ward identities into easily-analyzed polynomial equations.
\end{description}

The rest of the paper will then carry out proof of our main statement.  The steps are as follows:
\begin{description}
\item[In section \ref{charge-conservation-identities},]we will solve the Ward identity arising from
the action of the charge $Q_s$ arising from a spin $s$ current $j_s$ on the correlator 
$\vev{j_2j_2j_s}$ in the lightcone limit, where $j_2$ is the stress tensor.  We will show
that the only possible solution is given by the free-field solution.  This implies the existence of
infinitely many conserved currents of arbitrarily high spin,\footnote{The fact that the existence of
a higher-spin current implies the existence of infinitely many other higher-spin currents has been
proven before in \cite{Komargodski:2012ek} under the additional assumptions that the theory flows to a theory with a well defined S-matrix in the infrared, that the correlation function
$\vev{j_2j_2j_s} \neq 0$, and that the scattering amplitudes of the theory
have a certain scaling behavior.  This statement was also proven in \cite{Boulanger:2013zza} by
classifying all the higher-spin algebras in four dimensions.  We give a proof for the sake of
completeness, and also because our techniques differ from those two papers.} thereby giving rise to infinitely many charge conservation laws which powerfully constrain the theory.
\item[In section \ref{bilocal-fields},]we will construct certain quasi-bilocal fields which roughly 
behave like products of free fields in the lightcone limit, yet are defined for any CFT.  We will
establish that all the higher-spin charges (whose existence was proven in the previous step) 
act on these quasi-bilocals in a particularly simple way.
\item[In section \ref{qb-correlators},]we will translate the action of the higher-spin charges on
the quasi-bilocals into constraints on correlation functions of the quasi-bilocals.  We will then
show that these constraints are so powerful that they totally fix every correlation function of the
quasi-bilocals to agree with the corresponding correlation function of a bilocal operator in a
free-field theory.  
\item[In section \ref{constraining-correlators},]we show how the quasi-bilocal correlation functions can
be used to prove that the three-point function of the stress tensor must be equal to the three-point
function of either the free boson, the free fermion, or the free vector field, even away from the
lightcone limit.  This is then used to recursively constrain every correlation function of the CFT
to be equal to the corresponding correlation function in the free theory, finishing the proof.
\end{description}
This strategy is similar to the argument in the three-dimensional case given in \cite{Maldacena:2011jn}.  
There are two main differences between the three-dimensional and four-dimensional case that we must
account for:

First, the four-dimensional Lorentz group admits asymmetric representations, but the
three-dimensional Lorentz group does not.  By asymmetric, we mean that a current
$J_{\mu_1\dots\mu_n}$ is not invariant with respect to interchange of its indices.  In the standard
$(j_1,j_2)$ classification of representations of the Lorentz group induced from the isomorphism of
Lie algebras $\mathfrak{so}(3,1)_\bC \cong \mathfrak{sl}(2,\bC) \oplus \mathfrak{sl}(2,\bC)$, these are
the representations with $j_1 \neq j_2$.  The existence of these representations means that many more
structures are possible in four dimensions than in three (the asymmetric structures), and so many
more coefficients have to be constrained in order to solve the Ward identities.  We restrict
our attention to Ward identities arising from the action of a symmetric charge to correlation 
functions of only symmetric currents; we will then show that asymmetric structures cannot appear in
these Ward identities, making the exact solution of the identities possible.

Second, the space of possible correlation functions consistent with conformal symmetry is larger in
four dimensions than in three.  For example, consider the three-point function of the stress tensor
$\vev{j_2j_2j_2}$.  It has long been known (see,
e.g.~\cite{Stanev:1988ft}\cite{Osborn:1993cr}\cite{Stanev:2012nq}\cite{Zhiboedov:2012bm}) that this correlation function
factorizes into three structures in four dimensions, as opposed to only two structures in three
dimensions (ignoring a parity-violating structure which is eliminated in three dimensions by the
higher-spin symmetry).  These three structures correspond to the correlation functions that appear
in the theories of free bosons, free fermions, and free vector fields.  We will show that even
though more structures are possible in four dimensions, the Ward identities we need can still be
solved.

{\bf Note:} While this paper was being prepared, a paper by Stanev \cite{Stanev:2013qra} appeared, in
which the four, five, and six-point correlation functions of the stress tensor were
constrained in CFT's with a higher spin current in four dimensions.  It was also shown that the pole
structure of the general $n$-point function of the stress tensor coincides with that of a free field
theory.  Though this paper reaches the same conclusions as that paper, we do not make the
rationality assumption \cite{Nikolov:2000pm} of that paper.
\section{Definition of the lightcone limits}\label{lightcone-limits} 
The fundamental technical tool we need to extend into four dimensions is the \textit{lightcone
limit}.  In order to constrain the correlation functions of the theory to be equal to free field
correlators, we will show that the three-point function of the $\vev{j_2j_2j_2}$ must be equal to
$\vev{j_2j_2j_2}$ for a free boson, a free fermion, or a free vector - it cannot be some linear
combination of these three structures. To this end, it will be helpful to split up the Ward identities
of the theory into three different identities, each of which involves only one of the three
structures separately.  To do this, we will need to somehow project all the three-point functions of
the theory into these three sectors.  The lightcone limits accomplish this task.

Before defining the lightcone limits, we will set up some notation.  As in \cite{Maldacena:2011jn}, 
we are writing the flat space metric $ds^2 = dx^+dx^- + d\vec{y}^2$ and contracting each current
with lightline polarization vectors whose only nonzero component is in the minus direction: $j_s
\equiv J_{\mu_1\dots\mu_s}\epsilon^{\mu_1}\dots\epsilon^{\mu_s} = J_{--\dots-}$.  We will also
denote $\pd_1 \equiv \pd/\pd x_1^-$ and similarly for $\pd_2$ and $\pd_3$.  Thus, in all
expressions where indices are suppressed, those indices are taken to be minus indices.
There are two things we will establish:
\begin{enumerate}
\item We need to define an appropriate limit for each of the three cases, which, when applied to a
three-point function of conserved currents $\vev{\underline{j_{s_1}j_{s_2}}j_{s_3}}$, yields an
expression proportional to an appropriate correlator of the free field theory. For example, in the
bosonic case where all the currents are symmetric, we would like the lightcone limit to give us
$\pd_1^{s_1}\pd_2^{s_2}\vev{\phi\phi^*j_{s_3}}_{\text{free}}$.
\item Second, we need to explicitly compute the free field correlator which we obtain from the
lightcone limits. In the bosonic case where all currents are symmetric, this would mean that we
need to compute the three-point function $\vev{\phi\phi^*j_{s_3}}$ in the free theory.
\end{enumerate}

For the first task, we claim that the desired lightcone limits are:
\begin{align}
\vev{\underline{j_{s_1} j_{s_2}}_b j_{s_3}} &\equiv \lim_{|y_{12}|\rightarrow 0} |y_{12}|^2 
\lim_{x^+_{12}\rightarrow 0} \vev{j_{s_1} j_{s_2} j_{s_3}} \propto
\pd_1^{s_1}\pd_2^{s_2}\vev{\phi\phi^*j_{s_3}}_{\text{free}} \label{e:bosonic-lcl}\\
\vev{\underline{j_{s_1} j_{s_2}}_f j_{s_3}} &\equiv \lim_{|y_{12}|\rightarrow 0} |y_{12}|^4
\lim_{x^+_{12}\rightarrow 0} \frac{1}{x^+_{12}} \vev{j_{s_1} j_{s_2} j_{s_3}} \propto
\pd_1^{s_1-1}\pd_2^{s_2-1}\vev{\psi\C_-\bar{\psi}j_{s_3}}_{\text{free}} \label{e:fermionic-lcl}\\
\vev{\underline{j_{s_1} j_{s_2}}_v j_{s_3}} &\equiv \lim_{|y_{12}|\rightarrow 0} |y_{12}|^6 
\lim_{x^+_{12}\rightarrow 0} \frac{1}{(x^{+}_{12})^2}\vev{j_{s_1} j_{s_2} j_{s_3}} \propto
\pd_1^{s_1-2}\pd_2^{s_2-2}\vev{F_{-\alpha}F_{-\alpha}j_{s_3}}_{\text{free}} \label{e:vector-lcl}
\end{align}
Here, the subscript \textit{b, f,} and \textit{v} denote the bosonic, fermionic, and vector
lightcone limits.  $\phi$ is a free boson, $\psi$ is a free fermion, and $F$ is the 
field tensor for a free vector field.  The justification for the first two equations comes from the 
generating functions obtained in \cite{Stanev:2012nq}\cite{Zhiboedov:2012bm}; in those references, the three-point
functions for correlation functions of conserved currents with $y_{12}$ and $x_{12}^+$ dependence of
those types was uniquely characterized, and so taking the limit of those expressions as indicated
gives us the claimed result. In the vector case, \cite{Zhiboedov:2012bm} did not find a unique
structure, but rather, a one-parameter family of possible structures. Nevertheless, all possible
structures actually coincide in the lightcone limit, as is proven in appendix
\ref{vector-uniqueness}.  

We note that parity-violating structures cannot appear after taking these lightcone limits.
This is because the all-minus component of every parity violating structure allowed by 
conformal invariance in four dimensions is identically zero, as is easily checked from the explicit
forms given in \cite{Stanev:2012nq}\footnote{A more direct argument that does not require
explicit calculation can be made.  All parity-violating structures for three-point functions
consistent with conformal symmetry must have exactly one $\epsilon_{\mu_1\mu_2\mu_3\mu_4}$ tensor
contracted with polarization vectors and differences in coordinates.  Only two of these differences
are independent of each other, and all polarization vectors in the all-minus components are set to
be equal.  Thus, there are only three unique objects that can be contracted with the $\epsilon$
tensor, but we need four unique objects to obtain a nonzero contraction.  Thus, all parity-violating
structures have all-minus components equal to zero.}

Now, we will compute the free field three-point functions. The computation for each of the three
cases is straightforward; we demonstrate the calculation explicitly only for the bosonic case. Our
goal is to explicitly compute $\vev{\phi\phi^*j_s}$ on the lightcone. 
Then, using the explicit forms of the currents \cite{Mikhailov:2002bp}:
\begin{align}
j_s &= \sum_{k=0}^s c_k\pd^k\phi\pd^{s-k}\phi^* \\
c_k &= \frac{(-1)^k}{k!\left(k+\frac{d-4}{2}\right)!(s-k)!\left(s-k+\frac{d-4}{2}\right)!}
\end{align}
We can compute $\vev{\phi(x_1)\phi^*(x_2)j_s(x_3)}$ directly using Wick's theorem:
\begin{align}
\nonumber \vev{\phi(x_1)\phi^*(x_2)j_s(x_3)} 
\nonumber &= \sum c_i (\pd_3^i\vev{\phi(x_1)\phi^*(x_3)}) (\pd_3^{s-i}\vev{\phi(x_3)\phi^*(x_2)})\\
\nonumber &= \sum c_i (\pd_1^i\vev{\phi(x_1)\phi^*(x_3)}) (\pd_2^{s-i}\vev{\phi(x_3)\phi^*(x_2)}),
\text{ by translation invariance}\\
\nonumber &\propto \sum c_i \pd_1^i\pd_2^{s-i}\frac{1}{(\hat{x}_{13}\hat{x}_{23})^{d-2/2}}\\
\nonumber &= \frac{(-1)^s}{s!(\frac{d-4}{2})!} \sum_{k=0}^s
\frac{(-1)^ks!}{k!(s-k)!}\frac{1}{\hat{x}_{13}^{k+(d-2)/2}\hat{x}_{23}^{s-k+(d-2)/2}}\\
&\propto \frac{1}{(\hat{x}_{13}\hat{x}_{23})^{d-2/2}}
\left(\frac{1}{\hat{x}_{13}}-\frac{1}{\hat{x}_{23}}\right)^s \label{e:free-boson-correlator}
\end{align}
Here, we have defined $\hat{x}_{13} = x_{13}^- + \frac{\vec{y}_{13}}{x_{13}^+}$ and similarly 
for $\hat{x}_{23}$.  We have omitted an overall factor of $(x_{13}^+)^{(2-d)/2}$ which is common to
all correlators and will not matter for our calculations.

The fermionic case proceeds in precisely the same way. The relevant results are tabulated below:
\begin{align}
j_s &=
\sum_{k=2}^{s-1}\frac{(-1)^k\binom{s-1}{k}\binom{s+d-3}{k+\frac{n}{2}-1}}{\binom{s+d-3}{\frac{n}{2}-1}}
\pd^k\bar{\psi}\gamma\pd^{s-k-1}\psi \\
\vev{\psi_1\bar{\psi_2}j_s} &\propto \frac{1}{(\hat{x}_{13}\hat{x}_{23})^{d-1/2}}
\left(\frac{1}{\hat{x}_{13}}-\frac{1}{\hat{x}_{23}}\right)^{s-1}
\end{align}
Here, note that we have suppressed the spinor indices of $\psi_1$ and $\psi_2$.  They are set to $1$
and $\dot{1}$, respectively, so that the corresponding expression in vector indices has all minus 
indices.\footnote{Recall $x_{1\dot{1}} = x_-$.  See appendix \ref{representation-theory} for more
details.}

For the vector case, let $c_{ab}$ and $\bar{c}_{\dot{a}\dot{b}}$ be the self-dual and anti-self-dual
parts of the field tensor - in spinor indices, we would write $F_{ab\dot{a}\dot{b}} =
\epsilon_{ab}\bar{c}_{\dot{a}\dot{b}} + \epsilon_{\dot{a}\dot{b}}c_{ab}$.  Then, again setting the
spinor indices $a,b = 1$ and $\dot{a}, \dot{b} = \dot{1}$ to keep the corresponding vector indices
in the minus direction, we have:
\begin{align}
j_s &=
\sum_{k=1}^{s-1}\frac{(-1)^k\binom{s-2}{k}\binom{s+d-2}{k+\frac{n}{2}}}{\binom{s+d-2}{\frac{n}{2}-1}}
\pd^k\bar c\pd^{s-k-2}\bar{c} \label{e:c-definition}\\
\vev{c_1\bar c_2j_s} &\propto \frac{1}{(\hat{x}_{13}\hat{x}_{23})^{d/2}}
\left(\frac{1}{\hat{x}_{13}}-\frac{1}{\hat{x}_{23}}\right)^{s-2}
\end{align}
where here, we have suppressed the spinor indices by defining $c = c_{11}$ and
$\bar{c}=\bar{c}_{\dot{1}\dot{1}}$.  Before continuing, we emphasize that these three limits do not
cover all possible lightcone behaviors which can be realized in a conformal field theory.  We define
only these three limits because one crucial step in our proof is to constrain the three-point
function of the stress tensor $\vev{j_2j_2j_2}$, which has only these three scaling behaviors.  

Furthermore, though we have discussed only symmetric currents, one could hope that similar
expressions could be generated for asymmetric currents - that is, lightcone limits of correlation
functions of asymmetric currents are generated by one of the three free field theories discussed
here.  Unfortunately, running the same argument in \cite{Zhiboedov:2012bm} fails in the case of
asymmetric currents in multiple ways.  Consider the current $\vev{j_2j_s\bar{j}_s}$, where $j_s$ is
some asymmetric current and $\bar{j}_s$ is its conjugate.  To determine how such a correlator could
behave the lightcone limit, one could write out all the allowed conformally invariant structures
consistent with the spin of the fields, and seeing how each one behaves in the lightcone limits.
Unlike the symmetric cases, one finds that in the lightcone limit many independent structures exist, 
and these structures behave differently depending on which pair of coordinates we take the
lightcone limit.  To put it another way, for a symmetric current $s$, one has the decomposition:
\[\vev{j_2j_sj_s} = \sum_{j \in \{b,f,v\}} \vev{j_2j_sj_s}_j  \]
where the subscript $j$ denotes the result after taking corresponding lightcone limit in any of
the three pairs of coordinates (all of which yield the same result), and the corresponding
structures can be understood as arising from some free theory.  In the case of asymmetric $j_s$,
this instead becomes a triple sum
\[\vev{j_2j_s\bar{j}_s} = \sum_{j,k,l \in \{b,f,v\}} \vev{j_2j_s\bar{j}_s}_{(j,k,l)}\]
where each sum corresponds to taking a lightcone limit in each of the three different pairs of
coordinates, and we do not know how to interpret the independent structures in terms of a free field
theory.  This tells us that for asymmetric currents, the lightcone limit no longer achieves its
original goal of helping us split up the Ward identities into three identities which can be analyzed
independently; each independent structure could affect multiple different Ward identities.
Again, we emphasize that this does not exclude the possibility of a different lightcone limit reducing
the correlators of asymmetric currents to those of some other free theory.  It simply means that our
techniques are not sufficient to constrain correlation functions involving asymmetric currents, so
we will restrict our attention to correlation functions that involve only symmetric currents.

\section{Basic properties of form factors}\label{form-factors}
In this section, we will discuss how the spinor-helicity formalism and Fourier transformation allows
us to express correlation functions as simple polynomials.  This will make analysis of charge
conservation identities much easier.  From this section on, in order to make formulas easier to
read, we will often write $s$ instead of $j_s$ when referring to a current of spin $s$ in a
correlation funtion. For example, the three-point function of the stress tensor would be written as
$\vev{222}$.  As in the previous section, to simplify the discussion, we will begin with the bosonic
case, and comment on the fermionic and vector cases afterwards.

\subsection{Bosonic case}
We will consider the Fourier space expression of the spin $s$ current $j_s(p)$ of a free scalar theory
contracted with spinors $\ld_\A$ and $\tilde{\ld}_\da$.  For a symmetric spin $s$ current, for
example, this contraction is written explicitly in spinor indices\footnote{More details about the
representation theory of the Lorentz group in spinor indices can be found in appendix
\ref{representation-theory}.} as $\ld_{\A_1}\dots\ld_{\A_s}\tilde{\ld}_{\db_1}\dots\tilde{\ld}_{\db_s} 
j_s^{\A_1\dots\A_s\db_1\dots\db_s}$.  Since such a current is bilinear in the fields, we may consider 
the matrix element $F_s$ of this current in a two-particle state:
\begin{equation}
F_s \equiv \bra{p_1, p_2} \ld \cdots \ld \tilde{\ld} \cdots \tilde{\ld} j_s(p) \ket{0} \end{equation}
Note that here (and from now on), we suppress spinor indices, which are always contracted in the obvious 
way. First, we will establish some basic properties about the form factors $F_s$, and then we will
explain precisely how the $F_s$ are related to lightcone limits of three-point functions. 

Recall (from, e.g. \cite{Cachazo:2005ga}) that in four dimensions, some lightlike momentum $p_\m$
can be represented as a product of spinors as

\begin{equation}p_{\A\da} \equiv p_\m\s^\m_{\A\da} = \pi_\A\tilde{\pi}_\da\end{equation}
Lorentz invariance requires that $F_s$ be built from contractions of the $\pi$'s with the
$\lambda$'s (or otherwise vanish). However, there is a scaling symmetry given by:
\begin{equation}
(\pi,\tilde{\pi}) \rightarrow \left(z\pi, z^{-1}\tilde{\pi}\right), \,\,\,\,\, z \in \bC \backslash \{0\} 
\label{e:scaling-symmetry}
\end{equation}
 and both $\pi_i$ can be rotated independently. Enforcing this symmetry reduces the
possible structures that can appear to just $x\tilde{x}$ and $y\tilde{y}$, where
\begin{align*}
x &\equiv \ld\pi_1 & \tilde{x} &\equiv \tilde{\ld}\tilde{\pi}_1 
& y &\equiv \ld\pi_2 & \tilde{y} &\equiv \tilde{\ld}\tilde{\pi}_2 
\end{align*}
Counting spinors clearly forces $F_s$ to be homogeneous of degree $2s$ in these variables.
Then, letting $q = p_1+p_2$, we have that current conservation implies the equation
\begin{equation}q^{\A\db} \frac{\partial^2}{\partial \ld^{\alpha}\tilde{\ld}^{\db}} F = 0\end{equation}
A short computation verifies that this equation reduces to
\begin{equation}(\pi_1 \pi_2)(\tilde{\pi}_1 \tilde{\pi}_2) \left(\frac{\partial^2}{\partial x\partial \tilde{x}} + 
\frac{\partial^2}{\partial y\partial \tilde{y}}\right)F_s(x\tilde{x},y\tilde{y}) = 0\end{equation} 
The solutions to this equation are simple in the bosonic case, and we also independently derived
them from the explicit expressions for the currents found in \cite{Mikhailov:2002bp}:
\begin{equation}
F_s^{\text{boson}} = (y\tilde{y})^s{_2}F_1\left(-s,-s,1,-\frac{x\tilde{x}}{y\tilde{y}}\right),
\label{e:boson-hypergeo}
\end{equation}
where ${_2}F_1$ is the hypergeometric function. 

To see how $F_s$ relates to lightcone limits of three-point functions, note that this function is a
symmetric, homogenous degree $s$ polynomial in the variables $x\tilde{x}$ and $y\tilde{y}$. If we
explicitly expand the hypergeometric function, we see that the coefficient of the
$(x\tilde{x})^k(y\tilde{y})^{s-k}$ term in $F_s$ is equal to $c_k/c_0$, where $c_k$ are the
coefficients in the $\pd_1^k\phi\pd_2^{s-k}\phi^*$ term of $j_k$ given in section
\ref{lightcone-limits}. Thus, the form factors $F_s$ simply translate the differential structure of
$s_3$ into some polynomial in the spinor helicity variables in essentially the same way as the
Fourier transform.

To make this precise, recall that we established in the previous section that the lightcone limit of
three-point functions of CFT's with symmetric currents will agree with the result of the free
theory in the sense that:
\begin{equation}
\vev{\underline{s_1s_2}_bs_3} \xrightarrow{\text{lightcone limit}}
\partial_1^{s_1}\partial_2^{s_2}\vev{\phi\phi^*s_3}_{\text{free}}
\end{equation}
However, we have:
\begin{align*}
\vev{\phi\phi^*s_3}_{\text{free}} &= \sum c_i
\vev{\phi(x_1)\phi^*(x_2)\pd_3^i\phi^*(x_3)\pd_3^{s-i}\phi(x_3)}\\
&= \sum c_i \pd_3^i\vev{\phi(x_1)\phi^*(x_3)}\pd_3^{s-i}\vev{\phi(x_3)\phi^*(x_2)}\\
&= \sum c_i (-1)^s \pd_1^{s-i}\pd_2^i\vev{\phi(x_1)\phi^*(x_3)}\vev{\phi(x_3)\phi^*(x_2)}
\end{align*}
Immediately, it can be noted by taking $\pd_1 \rightarrow x\tilde{x}$ and $\pd_2 \rightarrow
y\tilde{y}$ that we would obtain exactly the form factor $F$ up to the overall $1/c_0$ constant. So
the structure is indeed well defined. This analysis flows exactly analogously for the fermion and
vector cases. However, for the sake of completeness, we note that we can really interpret the form
factors as being a Fourier transform of the correlation function and not just a formal
correspondence between differential operators and polynomials. The proof of this statement for each
of the three cases is in appendix \ref{fourier-transforms}.

\subsection{Fermionic case}\label{fermionic-case}
In this case, our external particle states come equipped with a helicity, which complicates the
analysis somewhat. To review, recall the chiral Dirac equation for a negative helicity spinor is
\begin{equation}
0 = i\s_{\A\da}^\m \pd_\m\psi^\A
\end{equation}
Our expected plane wave solution $\psi^\A = \mu^\A \exp(ip\cdot x)$ solves this equation only if
$p_{\A\da} \mu^\A = \pi_\A \tilde{\pi}_\da \mu^\A = 0$ - i.e. $\mu$ is proportional to $\pi$.
Analogous remarks hold for the spinor of positive helicity; its wavefunction brings a helicity
vector proportional to $\tilde{\pi}$. 

Now, suppose we are given a symmetric current $J_s$ with $q$ undotted and $q$ dotted indices.  We would
like to compute the matrix element of $J_s \ket{0}$ with some two-particle state.  The most general
structure that could appear has the form:
\begin{equation}
(\ld\pi_1)^a(\ld\pi_2)^b(\pi_1\pi_2)^c(\tilde{\ld}\tilde{\pi}_1)^d(\tilde{\ld}\tilde{\pi}_2)^e(\tilde{\pi}_1\tilde{\pi}_2)^f 
\end{equation}
Then we impose constraints on the coefficients as follows:
\begin{enumerate}
\item $J_s$ is spin $s$, so:
\begin{equation}a+b+d+e = s\end{equation}
\item $J_s$ is in the representation $\left(\frac{1}{2}q,\frac{1}{2}q\right)$, i.e. it has $q$ dotted and $q$ undotted indices, so:
\begin{equation}a+b = q\end{equation}
\begin{equation}d+e = q\end{equation}
\item $\pi_1$ and $\pi_2$ both have helicities $h_1$ and $h_2$, which can take on values $\pm 1/2$. 
This imposes the constraints:
\begin{equation}a+c-d-f= -2h_1\end{equation}
\begin{equation}b+c-e-f= -2h_2\end{equation}
\item The current must be conserved - i.e. the conformal dimension of the current has to be $d-2+s =
s+2$ (since $d=4$). Also, an $n$ particle state of definite momentum with particles of spin $j$ has
dimension $-2jn$, which is $-2$ in our case. All in all, $F$ should have dimension $s+2-2 = s$.
Since $E \sim p^2$ and $p \sim \pi \pi$, each contraction of spinors has dimension $1/2$. This
gives us:
\begin{equation}\frac{1}{2}\left(a+b+c+d+e+f\right) = s\end{equation}
\end{enumerate}
After imposing these constraints, the original six-parameter family is reduced to a one-parameter 
family of permissible terms - in particular, $h_1$ and $h_2$ must have opposite helicities.  To fix
the coefficients, we take an arbitrary linear combination of terms and demand that the sum be a
conformal primary (because $J$ is a conformal primary) - i.e. it should be invariant under special
conformal transformations. This constraint can be expressed as a simple differential equation in the
spinor-helicity language \cite{Witten:2003nn}: 
\begin{equation}
K_{\A\da}F =\left(\frac{\pd^2}{\pd\ld_1^\A\pd\tilde{\ld}_1^{\da}} + 
\frac{\pd^2}{\pd\ld_2^\A\pd\tilde{\ld}_2^{\da}}\right)F = 0 
\end{equation} 
This generates a recursion relation among the coefficients that can be solved explicitly.  As
before, let us define 
\begin{align*}
x &\equiv \ld\pi_1 & \tilde{x} &\equiv \tilde{\ld}\tilde{\pi}_1 
& y &\equiv \ld\pi_2 & \tilde{y} &\equiv \tilde{\ld}\tilde{\pi}_2 
\end{align*}
Then, the solution is
\begin{equation}
F^\text{fermion}_s \equiv \bra{0}J_s \ld^s \tilde{\ld}^s \ket{p_1-,p_2+} \propto 
x\tilde{y}(y\tilde{y})^{s-1}{_2}F_1\left(1-s,-s,2,-\frac{x\tilde{x}}{y\tilde{y}}\right)
\end{equation}
As a check, we note that these form factors could have been computed using explicit expressions 
for all the currents in the free fermionic theory, which can be found, e.g. in \cite{Anselmi:1999bb},
and our results agree. 

\subsection{Vector case}
In this case, our external particle states come with a polarization vector $\e_\m$ satisfying $\e
\cdot p = 0$ and a gauge symmetry $\e \rightarrow \e + wp$ for any constant $w$. Unlike in the
fermion case, there is no canonical choice for a polarization vector given $p$ and a choice of
helicity. Once we pick a decomposition $p_{\A\da} = \pi_{\A}\tilde{\pi}_{\da}$, the corresponding 
negative (resp.~positive) helicity polarization vector can be written in terms of an arbitrary
positive helicity spinor $\tilde{\m}$ (resp.~negative helicity spinor $\m$) according to
\begin{equation}\e_{\A\da} = \frac{\pi_\A\tilde{\m}_\da}{[\tilde{\pi}\tilde{\m}]}, \,\,\,\,\ 
\tilde{\e}_{\A\da} = \frac{\m_\A\tilde{\pi}_\da}{\vev{\pi\m}} \end{equation}
Fortunately, up to a gauge transformation $\tilde{\mu} \rightarrow \tilde{\mu} + 
\eta'\tilde{\pi}$, this polarization vector is independent of the choice of $\tilde{\mu}$. Furthermore,
under the scaling $(\pi,\tilde{\pi}) \rightarrow (z\pi,z^{-1}\tilde{\pi})$, $\epsilon_{\A\da}$
scales as $z^2$, which is what we expect for a helicity $-1$ particle.  Similar remarks hold for
$\tilde{\epsilon}_{\A\da}$.

Now, we can solve the system of linear equations determining $(a,b,c,d,e,f)$ as before, except now,
the helicities $h_i$ take on the values $\pm 1$ instead. Exactly the same analysis as before allows 
us to compute the relevant form factor:
\begin{equation}
F^{\text{vector}}_s \equiv \bra{0}J_s \ld^s \tilde{\ld}^s \ket{p_1-,p_2+} \propto 
(x\tilde{y})^2(y\tilde{y})^{s-2}{_2}F_1\left(2-s,-s,3,-\frac{x\tilde{x}}{y\tilde{y}}\right)
\end{equation}

\section{Charge conservation identities}\label{charge-conservation-identities}
We will now use the results of the previous section to prove that every CFT with a higher-spin
current contains infinitely many higher-spin currents of arbitrarily high (even) spin.  We will do this by
analyzing the constraints that conservation of the higher-spin charge imposes.  We treat the
bosonic, fermionic, and vector cases separately.

Before beginning, we will tabulate a few results about commutation relations that we will use freely
throughout from this section onwards.  Their proofs are identical to those in
\cite{Maldacena:2011jn}, and are therefore omitted:
\begin{enumerate}
\item If a current $j'$ appears (possibly with some number of derivatives) in the commutator
$[Q_s,j]$, then $j$ appears in $[Q_{s},j']$.  
\item Three-point functions of a current with odd spin with two identical currents of even spin are
zero: $\vev{j_sj_sj_{s'}} = 0$ if $s$ is even and $s'$ is odd.
\item The commutator of a symmetric current with a charge built from another symmetric current
contains only symmetric currents and their derivatives:
\begin{equation}
[Q_s,j_{s'}] = \sum_{s'' = \max[s'-s+1,0]}^{s'+s-1} \alpha_{s,s',s''}\pd^{s'+s-1-s''}j_{s''}
\label{e:commutator-expansion}
\end{equation}
The proof of this statement requires an additional step since one needs to exclude asymmetric
currents contracted with invariant symbols like the $\epsilon$ tensor.  For example, consider what
structures could appear in $[Q_2,j_2]$.  This object has three dotted and three undotted spinor
indices, so one could imagine that a structure like
$\epsilon_{ab}j^{abcde\dot{c}\dot{d}\dot{e}}$ could appear in $[Q_2,j_2]$.  However, $[Q_2,j_2]$ has
conformal dimension $5$, and the unitarity bound constrains the current $j$, which transforms in the
$(5/2,3/2)$ representation, to have conformal dimension at least $d-2+s = 6$, which is impossible.
The proof for a general commutator $[Q_s,j_{s'}]$ follows in an identical manner.
\item $[Q_s,j_2]$ contains $\pd j_s$.  This was actually proven for all dimensions in appendix A 
of \cite{Maldacena:2011jn}.  Item $1$ then implies that $[Q_s,j_s]$ contains $\pd^{2s-3}j_2$.
\end{enumerate}
In these statements, we are implicitly ignoring the possibility of parity violating structures.  For
example, the three-point function $\vev{221}$, which is related to the $U(1)$ gravitational anomaly, 
may not be zero in a parity violating theory.  As mentioned in section \ref{lightcone-limits},
however, the all-minus components of every parity-violating structure consistent with conformal
symmetry is identically zero, so they will not appear in any of our identities here.

Let's start with the bosonic case:

Consider the charge conservation identity arising from the action of $Q_s$ on $\vev{\underline{22}_bs}$:
\[0 = \vev{\underline{[Q_s,2]2}_bs} + \vev{\underline{2[Q_s,2]}_bs} + \vev{\underline{22}_b[Q_s,s]} \]
If $s$ is symmetric, we may use the general commutation relation (\ref{e:commutator-expansion}) and
the lightcone limit (\ref{e:bosonic-lcl}) to expand this equation out in terms of free field
correlators:
\begin{equation}
0 = \partial_1^2\partial_2^2\left(\C(\partial_1^{s-1}+(-1)^s\partial_2^{s-1})\vev{\underline{\phi\phi^*}s}_{free} +
\sum_{2\le k < 2s-1\text{ even}} \tilde{\alpha_k}\partial_3^{2s-1-k}\vev{\underline{\phi\phi^*}k}_{free}\right)
\label{e:bosonic-charge-identity}
\end{equation}
Note that the sum over $k$ is restricted to even currents since $\vev{22k} = 0$ for odd $k$.  In
addition, the fact that the coefficient in front of the $\pd^{s-1}_2$ term is constrained to be
$(-1)^s$ times the coefficient for the $\pd^{s-1}_1$ term arises from the symmetry of equation
(\ref{e:free-boson-correlator}) under interchange of $x_1$ and $x_2$.

Now, we apply our methods from section \ref{form-factors}. In the Fourier transformed spinor
variables, which turns derivatives into multiplication by the momenta, and therefore into
multiplication by the appropriate spinor products, we find that the action of the derivatives with
respect to each variables are $\partial_1 \sim x\tilde{x}$, $\partial_2 \sim y\tilde{y}$, and
$\partial_3 \sim x\tilde{x}+y\tilde{y}$. Furthermore, the action of $\phi$ and $\phi^*$ on the bra
$\bra{0}$ allows us to rewrite the charge conservation identity in terms of the matrix elements
$F_s$. After ``cancelling out'' the overall derivatives as before, the relevant equation is:
\begin{equation}
0 = \gamma((x\tilde{x})^{s-1}+(-1)^s(y\tilde{y})^{s-1})F_s(x\tilde{x},y\tilde{y}) +
\sum_{2\le k<2s-1\text{ even}} \tilde{\A}_k (x\tilde{x}+y\tilde{y})^{2s-1-k}
F_k(x\tilde{x},y\tilde{y})
\label{e:fourier-bosonic-charge-identity}
\end{equation}
The solution of (\ref{e:fourier-bosonic-charge-identity}) is not easy to obtain by direct calculation
in four dimensions.  We can make two helpful observations, however.  First, not all coefficients can
be zero.  This is because we know $2$ appears in $[Q_s,s]$, so at least $\tilde{\alpha}_2$ is not
zero.  Second, we know that the free boson exists (and is a CFT with higher spin symmetry), and
therefore, the coefficients one obtains from that theory would exactly solve this equation.  We will
show that this solution is unique.

Suppose we have two sets of coefficients $(\C,\{\tilde{\A}_k\})$ and $(\C',
\{\tilde{\B}_k\})$ that solve this equation. First, suppose $\C \neq 0$ and $\C' \neq 0$. Then, 
we can normalize the coefficients so that $\C = \C'$ are equal for the two
solutions. Then, subtract the two solutions from each other so that the $\C$ terms vanish. If we
evaluate the result at momenta such that $\tilde{x}=\tilde{y}$, we may absorb all overall $y$ and
$\tilde{y}$ factors into the coefficients and re-express the equation as a polynomial identity in a
single variable $z \equiv x/y$: 
\begin{equation}0 = \sum_{2\le k<2s-1\text{ even}} \tilde{\D}_k(1+z)^{2s-1-k}{_2}F_1(-k,-k,1,-z)\end{equation} 
Then, the entire right hand side is divisible by $1+z$ since $s$ is even, so we may divide both
sides by $1+z$. Setting $z=-1$, we find that ${_2}F_1(2-2s,2-2s,1,1) \neq 0$ implies
$\tilde{\delta}_{2s-2} = 0$. Then, the entire right hand side is proportional to $(1+z)^2$, so we
may divide it out. Then, setting $z=-1$ again, we find $\tilde{\delta}_{2s-4} = 0$. Repeating this
procedure, we conclude that all coefficients are zero, and therefore, that the two solutions are
identical. On the other hand, suppose one of the solutions has $\C = 0$. Then, the same argument
establishes that all the coefficients $\tilde{\A}_k$ are zero.  As noted earlier, however, the 
trivial solution is disallowed.  Therefore, the solution is unique and coincide with one for free
boson. Thus, we have infinitely many even conserved currents, as desired.

In the fermionic case, precisely the same analysis works.  The action of $Q_s$ on
$\vev{\underline{22}_fs}$ for symmetric $s$ leads to
\begin{equation}
0 = \partial_1^2\partial_2^2 \Big(\C(\partial_1^{s-2}+(-1)^{s-1}\partial_2^{s-2})\vev{\psi\bar\psi s} +
\sum_{2\le k < 2s-2\text{ even}} \tilde{\alpha}^k\partial_3^{2s-2-k}\vev{\psi\bar\psi k}\Big),
\end{equation}
Then, converting this expression to form factors and running the same analysis from the
bosonic case verbatim (i.e.~ take $\tilde{x}=\tilde{y}$, absorb factors of $\tilde{y}$ into the 
coefficients and write it as a polynomial in $z = -x/y$, etc.) establishes that the unique solution
to this equation is the one arising in the theory of a free fermion.

In the vector case, the argument again passes through exactly as before, except for one problem:
unlike in the bosonic and fermionic case, we do not have unique expressions for the
three-point functions of currents with the vector-type coordinate dependence, so this only
demonstrates that the free-field solution is an admissible solution, but not necessarily the unique
solution. Nevertheless, in the lightcone limit, all possible structures for
three-point functions coincide with the free-field answer.\footnote{Actually, we proved that
correlators of the form $\vev{22s}$ have a unique vector structure even away from the lightcone
limit. The proof, however, is very technical, and it is given in appendix \ref{uniqueness-of-22s}.}
This was proven in appendix \ref{vector-uniqueness}.

\section{Quasi-bilocal fields: basic properties}\label{bilocal-fields}
In this section, we will define a set of quasi-bilocal operators, one for each of the three
lightcone limits, and characterize their commutation relations with the conserved charges of the 
higher-spin currents.  These quasi-bilocals will turn out to mimic true bilocal products of free
fields in the lightcone limit, ultimately enabling us to prove that the three-point function of the
stress tensor can exhibit only one of the three possible structures allowed by conformal symmetry.
As in the three-dimensional case, we define the quasi-bilocal operators on the lightcone as operator
product expansions of the stress tensor with derivatives ``integrated out'':
\begin{align*}
\underline{22}_b&=\partial_1^2\partial_2^2B(\underline{x_1,x_2}) \\
\underline{22}_f&=\partial_1\partial_2F_{-}(\underline{x_1,x_2}) \\
\underline{22}_v&=V_{--}(\underline{x_1,x_2})
\end{align*}
The motivation behind these definitions can be understood by appealing to what these expressions
look like in free field theory.  There, they will transform like simple bilocal products of free
fields:
\begin{align*}
B(x_1,x_2)&\sim :\phi(x_1)\bar \phi(x_2):+:\phi(x_2)\bar\phi(x_1):
\\
F_{-}(x_1,x_2)&\sim :\psi_{1}(x_1)\bar\psi_{\dot 1}(x_2):-:\psi_{1}(x_2)\bar\psi_{\dot 1}(x_1):
\\
V_{--}&\sim :F_{-\alpha}(x_1)F_{-\alpha}(x_2):
\end{align*}
Also, it is clear from the basic properties of our lightcone limits that when they are inserted into
correlation functions with another conserved current $j_s$, they will be proportional to an
appropriate free field correlator.  Since $\vev{\underline{22}s} = 0$ for odd $s$, only the
correlation functions with even $s$ will be nonzero:
\begin{align*}
\vev{B(\underline{x_1,x_2})j_s} &\propto \vev{\phi(x_1)\phi(x_2)j_s(x_3)}_{\text{free}}\\
\vev{F_-(\underline{x_1,x_2})j_s} &\propto \vev{\psi(x_1)\bar{\psi}(x_2)j_s(x_3)}_{\text{free}}\\
\vev{V_{--}(\underline{x_1,x_2})j_s} &\propto \vev{c(x_1)\bar{c}(x_2)j_s(x_3)}_{\text{free}}
\end{align*}
Of course, away from the lightcone, things will not be so simple: we have not even defined the 
quasi-bilocal operators there, and their behavior there is the reason why they are not true
bilocals.  These complications, however, will be important only in section \ref{qb-correlators}.
For now, the lightcone behavior alone is enough to establish the commutator of $Q_s$ with the
bilocals.  As usual, we begin with the bosonic case:

Assume that $\vev{\underline{22}_b2} \neq 0$. We claim that 
\begin{equation} 
[Q_s, B(\underline{x_1,x_2})] = (\partial^{s-1}_1+\partial^{s-1}_2) B(\underline{x_1,x_2}). 
\label{e:boson-bilocal-commutator}
\end{equation}
This can be shown using the same arguments as \cite{Maldacena:2011jn}.  To begin, notice that the
action of $Q_s$ commutes with the lightcone limit.  Thus,
\[\vev{[Q_s,B]j_k} = \vev{\underline{[Q_s,j_2]j_2}j_k} + \vev{\underline{j_2[Q,j_2]}j_k} 
= -\vev{\underline{j_2j_2}[Q_s,j_k]} = \vev{[Q_s,\underline{j_2j_2}]j_k}\]
This immediately leads to:
\begin{equation}
[Q_s,B(\underline{x_1,x_2})] = (\partial^{s-1}_1+\partial^{s-1}_2) \tilde B(\underline{x_1,x_2}) 
+ (\partial^{s-1}_1-\partial^{s-1}_2) B'(\underline{x_1,x_2}),
\end{equation}
Here, $\tilde B$ is built from even currents (of the free theory), while $B'$ is built from odd
currents. This makes the whole expression symmetric.  We would like to show that $B'=0$. Therefore,
suppose otherwise so that some current $j_{s'}$ has nontrivial overlap with $B'$. Then, the charge
conservation identity $0 = \vev{\left[Q_{s'},B' j_2\right]}$ yields
\begin{align}
0&=\vev{\left[ Q_{s'},B'(\underline{x_1,x_2})\right] j_2}+\vev{B'(\underline{x_1,x_2})\left[ Q_{s'},j_2\right]},
\\
\Rightarrow 0 &=\gamma \left(\partial^{s'-1}_1-\partial^{s'-1}_2\right) \vev{\underline{\phi \bar\phi} j_2} +\sum \limits_{k=0}^{s'+1}\tilde \alpha _k \partial^{s'+1-k}\vev{\underline{\phi\bar\phi}j_k}.
\end{align}
Using the same techniques as the previous section, we obtain 
\begin{equation}
0=\gamma \left( \left(x \tilde x \right)^{s'-1}-\left(y\tilde y \right)^{s'-1}\right) F_2\left( x\tilde x, y \tilde y\right)+\sum_{k=0}^{s'+1}\tilde\alpha_{k} \left(x\tilde x+y\tilde y\right)^{s'+1-k}F_{k}\left(x\tilde x, y\tilde y\right).
\end{equation}
In this sum, $\tilde\alpha_{s'}\neq0$ because $j_{s'} \subset \left[ Q_{s'},2\right]$.
Therefore, we can use the same procedure as before to show that all $\tilde\alpha_k$ are nonzero
if they are nonzero for the free field theory. In particular, since $\tilde{\alpha}_1$ is not 
zero for the complex free boson, the overlap between $j_1$ and $B'$ is not zero. Now, let's
consider 
\begin{equation}
0=\vev{\left[Q_s,Bj_1\right]}= \left( \partial_1^{s-1}-\partial_2^{s-1}\right)\vev{B'j_1}+\vev{B\left[Q_s,j_1\right]},
\end{equation}
where $Q_s$ is a charge corresponding to any even higher-spin current appearing in the operator product
expansion of $\underline{j_2j_2}_b$.  We have shown the first term is not zero.  We will prove that
the second term must be equal to zero to get a contradiction.  Specifically, we will show that there
are no even currents in $[Q_s,j_1]$.  Since $B$ is proportional to $\underline{22}$, and since
$\vev{22s} = 0$ for all odd s, this yields the desired conclusion.

Consider the action of $Q_s$ on $\vev{221}$.  After running the usual series of tricks, we have
\begin{equation}
0=\gamma \left( \left(x \tilde x \right)^{s-1}-\left(y\tilde y \right)^{s-1}\right) F_1\left(
x\tilde x, y \tilde y\right)+\sum_{k=0}^s\tilde\alpha_{k} \left(x\tilde x+y\tilde y\right)^{s-k}F_{k}\left(x\tilde x, y\tilde y\right).
\label{e:221-identity}
\end{equation}
We want to show that $\alpha_k = 0$ for even $k$.  Recall the definition of $F_k$
from equation (\ref{e:boson-hypergeo}):
\begin{align*}
F_k &= (y\tilde{y})^k{_2}F_1\left(-k,-k,1,-\frac{x\tilde{x}}{y\tilde{y}}\right)\\
&= \sum_{i=0}^k c^k_i (x\tilde{x})^i(y\tilde(y))^{k-i}
\end{align*}
The hypergeometric coefficients $c^k_i$ have the property that $c^k_i = (-1)^kc^k_{k-i}$.  
Now, we collect terms in equation (\ref{e:221-identity}) proportional to
$(x\tilde{x})^s$ and $(y\tilde{y})^s$ - each sum must vanish separately for the entire polynomial to
vanish.  We obtain
\begin{align*}
\gamma + \sum_{0 \le k \le s \text{  odd}}\alpha_ku_k + \sum_{0 \le k \le s \text{  even}}\alpha_kv_k 
= 0 \\
- \gamma - \sum_{0 \le k \le s \text{  odd}}\alpha_ku_k + \sum_{0 \le k \le s \text{  even}}\alpha_kv_k
= 0
\end{align*}
Here, $u_k$ and $v_k$ are sums of products of coefficients of the hypergeometric function
and the binomial expansion of $(x\tilde{x} + y\tilde{y})^{s-k}$; we do not care about their
properties except that, with the signs indicated above, they are strictly positive, as can be
verified by direct calculation.  By adding and subtracting these equations, we obtain two
separate equations that must be satisfied by the odd and even coefficients separately
\begin{align*}
\gamma + \sum_{0 \le k \le s \text{  odd}}\alpha_ku_k = 0 \\
\sum_{0 \le k \le s \text{  even}}\alpha_kv_k = 0
\end{align*}
Exactly analogously, we may do the same procedure to every other pair of monomials
$(x\tilde{x})^a(y\tilde{y})^{s-a}$ and $(x\tilde{x})^{s-a}(y\tilde{y})^a$ to turn the constraints
for the two monomials into constraints for the even and odd coefficients (where we're considering
$\gamma$ as an odd coefficient) separately.  Hence, by multiplying each term by the monomial from
which it was computed and then resumming, we find that the original identity (\ref{e:221-identity})
actually splits into two separate identities that must be satisfied.  For the even terms, this
identity is:
\[0 = \sum_{0 \le k \le s \text{  even}} \alpha_k(x\tilde{x} + y\tilde{y})^{s-k} (y\tilde{y})^k
{_2}F_1\left(-k,-k,1,-\frac{x\tilde{x}}{y\tilde{y}}\right) \]
Then, we may again use the tricks from section \ref{charge-conservation-identities} and conclude
that all $\alpha_k = 0$ for even $k$, which is what we wanted. Thus, $B'=0$.

Now we would like to show that $B=\tilde B$. First of all we will show that $\tilde B$ is nonzero.
Consider the charge conservation identity 
\[0 = \vev{\left[Q_s, B j_2\right]} = \left(\pd^{s-1}_1 + \pd^{s-1}_2\right)\vev{\tilde{B}2} +
\vev{B,[Q_s,2]}\]
Since $\left[Q_s, j_2\right]\supset\partial j_s$, and since $\vev{Bs} \neq 0$, the second term in
that identity is nonzero, and so $\tilde B$ must be nonzero. Now we can
normalize the currents in such a way that $j_2$ has the same overlap with $\tilde B$ and $B$. After
normalization, we know that $B-\tilde B$ does not contain any spin $2$ current because the stress
tensor is unique, by hypothesis. Now, we will show that $B-\tilde B$ is zero by contradiction.
Suppose $B-\tilde B$ is nonzero.  Then, there is a current $j_s$ whose overlap with $B-\tilde B$ is 
nonzero. Then, the charge conservation identity for the case $s>2$
is 
\begin{align}
0&=\vev{\left[Q_s,\left( B-\tilde B\right) j_2\right]},\\
0&=\gamma \left( \left( x\tilde x\right)^{s-1}+\left(y\tilde y\right)^{s-1}\right)F_2(x\tilde x, y\tilde y)+ 
\sum\limits_{k=0}^{s+1}\tilde \alpha_k (x\tilde x+y\tilde y)^{s+1-k} F_{k}\left( x\tilde x, y \tilde y\right),
\end{align}
where we assume that $\tilde\alpha_s\neq 0$. Then, we can again run the same analysis as section
\ref{charge-conservation-identities} to conclude
that since $\tilde \alpha_s\neq 0$, we must have $\tilde \alpha_2\neq 0$ - that is, $j_2$ has
nonzero overlap with $B-\tilde B$, which is a contradiction. It means that $B-\tilde B$ has no
overlap with any currents $j_s$ for $s>2$. The only possibility is to overlap only with spin zero
currents. Suppose that there is a current $j'_0$ that overlaps with $B-\tilde{B}$, where the prime
distinguishes it from a spin $0$ current $j_0$ that could appear in $B$.  We first show that
$\vev{j_0j_0'} = 0$.  Consider the charge conservation identity the action $Q_4$ on $\vev{(B-\tilde B)j_0}$.
The action of the charge is $\left[Q_4,0\right]=\partial^3 j_0+\partial j_2+\dots$, where the
$\dots$ represent terms that cannot overlap with $\underline{22}$ (from which $B$ is constructed) or
the even currents that appear in $\tilde{B}$. By hypothesis, $B-\tilde B$ has no overlap with $j_2$,
so the identity simplifies to $\vev{j_0j_0'} = 0$. Then, since $j_0'$ is nonzero, it should have
nontrivial overlap with some $Q_s$.  Now, recall the fact that if a current $j'$ appears (possibly
with some number of derivatives) in the commutator of $[Q_s,j]$, then $j$ appears in $[Q_s,j']$.
Thus, there should be a current current of spin $s''<s$ such that $\left
[Q_s,j_{s''}\right]=j_0'+\dots$. The action $Q_s$ on $\vev{\left(B-\tilde B \right)j_{s''}}$ is 

\begin{equation}
 \vev{\left[ Q_s,\left(B-\tilde B \right)j_{s''}\right]}=\partial^3_3\vev{\left(B-\tilde B \right)j_0'} +\partial\vev{\left(B-\tilde B\right) j_2},
\end{equation}
Here, we have used that the action of $Q_s$ on $B$ and $\tilde{B}$ is identical because $B' = 0$.
Then, since the second term is zero, thus the first term is equal to zero as well. Thus, $B-\tilde
B$ has no overlap with any currents and is equal to zero, as desired.

In the fermionic case, we can run almost the same argument as in the bosonic case, except there is no
discussion of a possible $j_0$, since there is no conserved spin zero current in the free fermion
theory. We obtain the action of the charge on the fermionic quasi-bilocal is 
\begin{equation}
 [Q_s, F_{-}(\underline{x_1,x_2})] = (\partial^{s-1}_1+\partial^{s-1}_2) F_{-}(\underline{x_1,x_2}). 
\end{equation}

In the vector case, we again can repeat the argument to obtain
\begin{equation}
 [Q_s, V_{--}(\underline{x_1,x_2})] = (\partial^{s-1}_1+\partial^{s-1}_2) V_{--}(\underline{x_1,x_2}), 
 \end{equation}
However, in this case we can consider $j_3$ instead of $j_1$, and we again do not have to consider $j_0$.

\section{Quasi-bilocal fields: correlation functions}\label{qb-correlators}
In this section, we will discuss how to sensibly extend the quasi-bilocal fields away from the
lightcone.  This will allow us to impose constraints on the correlation functions of the bilocals
from the full conformal group, not just the subgroup that keeps the arguments of the quasi-bilocals
on the lightcone.  With the commutation relations developed in the previous section, we will show
that this totally fixes all the correlation functions of the quasi-bilocal operators.

Let's begin by tackling the first problem of extending the quasi-bilocals away from the lightcone in a
sensible fashion.  Recall that the bilocals are operator product expansions of the stress tensor
with itself.  We know that this expansion will contain all the even higher-spin
currents and their descendants.  It may contain other kinds of operators, but the lightcone limit
is defined so that these operators are projected out.  This suggests that we can extend the
quasi-bilocals away from the lightcone limit by taking a general combination of conserved currents
and their descendants and enforcing that their correlation functions should behave like a true
bilocal in the lightcone limit.  For example, in the bosonic case, we may define
\begin{align}
B(x_1,x_2) &= \sum_{\text{even } s\ge 2} b_s(x_1,x_2) \\
b_s(x_1,x_2) &= \sum_{k,l} c_{kl}(x_1-x_2)^k\pd^lj_s\left(\frac{x_1+x_2}{2}\right)
\label{e:decomposition}
\end{align}
The coefficients $c_{kl}$ are totally fixed because in the lightcone limit,
$\vev{B(\underline{x_1,x_2})j_s} \propto \vev{\phi(x_1)\phi^*(x_2)j_s}$.  Since $\vev{j_sj_{s'}}
\propto \delta_{ss'}$ for conserved currents, only the spin $s$ sector could give the required limit
for a particular $j_s$.  That is, we must demand $\vev{b_s(\underline{x_1,x_2})j_s} \propto
\vev{\phi(x_1)\phi^*(x_2)j_s(x_3)}$ for each $b_s$ separately.  This relation fixes all the $c_{kl}$
for each $b_s$, and makes $b_s$ transform like a product of two free fields.  By comparing conformal
dimensions of both sides of (\ref{e:decomposition}), we find that the sum over $k$ and $l$ is 
then restricted to be over values of $k$ and $l$ such that $s+l-k = 0$.
After all these coefficients are fixed, there is no obstruction to taking $x_1$ and $x_2$ away from
the lightcone.  This construction works the same way for the fermionic and vector
quasi-bilocals with analogous results, except that the quasi-bilocals in those cases carry some spin
structure.

Evidently, under this definition the bosonic quasi-bilocal is a bi-primary field with a conformal
dimension of $1$ with respect to each argument.  It has the transformation properties of a
product of two free bosons since we required this of each $b_s$.  Put another way, each $b_s$ is like a
conformal block, giving the contribution of the spin $s$ primary and its descendants to the OPE of
two free fields.  This qualitative picture can be made sharp by noticing the two point function of
any particular $b_s$ actually does encode the contribution of the spin $s$ current to the four-point
of free fields, which is precisely what a conformal block is.  This answer in four dimensions is
known, and given in equation (2.11) of \cite{Dolan:2000ut}.  One can then check that for any
particular $s$, the scaling behavior of this solution near one of its singularities is logarithmic,
unlike the power-like singularities we would expect from a local operator.  Thus, each $b_s$ is 
not truly bilocal.  Nevertheless, we will now show that $B$, the full sum of all the $b_s$, is bilocal.

Let's consider an $n$-point function of quasi-bilocals.  In the bosonic case, the commutation
relation (\ref{e:boson-bilocal-commutator}) imposes the simple relation
\[\sum_{i=1}^{2n} \pd^{s-1}_i \vev{B(\underline{x_1,x_2})\dots B(\underline{x_{2n-1}x_{2n}})}, 
\,\,\,\,\,\, \text{for all even s} \]
As shown in appendix E of \cite{Maldacena:2011jn}, this fixes the $x^-$ dependence of the 
$n$-point function to have the particular form: 
\[\sum_{\sigma \in S^{2n}} g_{\sigma}\left(x^-_{\sigma(1)} - x^-_{\sigma(2)}, x^-_{\sigma(3)} -
x^-_{\sigma(4)},\dots, x^-_{\sigma(2n-1)}-x^-_{\sigma(2n)}\right) \]
where $S^{2n}$ is the set of permutations of $2n$ elements.  The point is that the $x_i^-$
dependence of the $n$-point function is constrained such that, for each $g_\sigma$, $x_i^-$ 
can only appear in a difference with one and only one other coordinate.  This is
a very strong constraint: each $g_\sigma$ in the above series can be written as a product of a
dimensionful function of distances that matches the conformal dimension of the bilocals and a
dimensionless function $\mathcal{F}(u,v)$ of conformal cross-ratios
$u = \frac{x_{12}^2x_{34}^2}{x_{13}^2x_{24}^2}$ and 
$v = \frac{x_{14}^2x_{23}^2}{x_{13}^2x_{24}^2}$.  The constraint on the functional form of
$g_\sigma$, however, forbids any $\mathcal{F}(u,v)$ except for the trivial function
$\mathcal{F}(u,v) = 1$, because $u$ and $v$ each separately violate the constraint.
Furthermore, by rotational invariance, translational invariance, and conformal covariance, 
the $n$-point function of bilinears should have the correct spin structure and conformal dimension
for a product of two free fields: it can only depend on distances between coordinates $d_{ij}$ and
have conformal dimension $1$ with respect to each variable.  Since $x_1$ and $x_2$ are lightlike
separated, $d_{12}$ cannot appear, and similarly for every pair of arguments of the same bilocal.
Since $B$ is symmetric in its two arguments, and since the $n$ point function must be symmetric
under interchange of any pair of the identical $B$'s, the form of the $n$-point function has to be
proportional to a sum of terms with equal coefficients, each of which is a product $\prod
d_{ij}^{-2}$, where the product has $n$ terms corresponding to some partition of the $2n$ points
into pairs where no pair contains two arguments of the same bilocal.  For example, the two-point
function is 
$$\tilde{N}\left(\frac{1}{d_{13}^{2}d_{24}^{2}} + \frac{1}{d_{14}^{2}d_{23}^{2}}\right),$$ 
where $\tilde{N}$ is a constant of proportionality.  One
immediately notes that these answers are proportional to the $n$-point function of
$:\phi(x_1)\phi(x_2):$ in a theory of free bosons.  

These results all extend to the fermionic and vector cases, with the sole difference that we need
to take into account the spin of the operators.  For example, in the fermionic case, instead of
having products built out of terms like $d_{ij}^{-2}$, we will instead have building blocks of
$x_{ij}^+d_{ij}^{-3}$, which yields the two-point function
$$\tilde{N}\left(\frac{x_{13}^+x_{24}^+}{d_{13}^{3}d_{24}^{3} }+ \frac{x_{14}^+x_{23}^+}{d_{14}^{3}d_{23}^{3}}\right)$$.  
In the vector case, the correct building blocks are $(x_{ij}^+)^2d_{ij}^{-4}$, which results in the
two-point function
$$\tilde{N}\left(\frac{(x_{13}^+)^2(x_{24}^+)^2}{d_{13}^{4}d_{24}^{4}} +
\frac{(x_{14}^+)^2(x_{23}^+)^2}{d_{14}^{4}d_{23}^{4}}\right)$$.

Now, let's fix the the overall constants in front of each $n$-point function.  We claim that they
all are fixed by the normalization of the two-point function of the bilocals.  This can be seen by
considering how one can obtain the $n$-point function of quasi-bilocals from the $n-1$ point
function.  We know the $n$-point function of some quasi-bilocal $\mathcal{A}$ is:
\[\underbrace{\vev{\mathcal{A}\dots\mathcal{A}}}_{n \text{ copies of } \mathcal{A}} = \tilde{N}_ng(d_{ij}) \]
where $g$ is some known function which agrees with the result for the
$n$-point function of the corresponding free theory bilocal.  Each bilocal contains the stress
tensor $j_2$ in its OPE, so we can consider acting on both sides with some projector $P$ which
isolates the contribution of $j_2$ from the first bilocal. We will see in the section 
\ref{constraining-correlators}, for example, that for the vector bilocal, this projector just sets
$x_1 = x_2$.  Then, we can integrate over the coordinate $x_1$.  This yields the action of the
dilatation operator on the $n-1$ point function, whose eigenvalue will be some multiple of the
conformal dimension of the free field.  So by this procedure, we can fix the coefficient in front of
the $n$-point function in terms of the $n-1$ point function.  So by recursion, all the coefficients
of the correlation functions are fixed by the coefficient $\tilde{N}$ appearing in front of the 
two-point function.

\section{Constraining all the correlation functions}\label{constraining-correlators}
We have shown now that the $n$-point functions of each quasi-bilocal field exactly coincides with
the result for a theory of $N$ free fields for some well-defined integer $n$.  Now, we will explain
how to use this fact to extract all the other correlation functions of the theory and fix them to be
equal to the free field result.  We will start by proving that the three point function $\vev{222}$
must be exactly equal to the result for a free boson, a free fermion, or a free vector, but not a 
combination of those structures.  That is, if we write the most general possible form:
\begin{equation}
\vev{222} = c_b\vev{222}_{\text{free boson}} + c_f\vev{222}_{\text{free fermion}} 
+ c_v\vev{222}_{\text{free vector}}
\end{equation}
The result will be consistent with the higher-spin symmetry only if $(c_b,c_f,c_v) \propto (1,0,0)$
or $(0,1,0)$ or $(0,0,1)$.  Since the lightcone limits project out each sector separately, it is
sufficient to show that if the bosonic lightcone limit $\vev{\underline{22}_b2}$ is nonzero, then the 
other two lightcone limits are zero, and similarly for the other two cases.

We first show that if $\vev{\underline{22}_b2} \neq 0$ then $\vev{\underline{22}_f 2} = 0 =
\vev{\underline{22_v}2}$. Consider the action of $Q_4$ on $\vev{\underline{22}_b2}$.  By exactly the
same analysis as the charge conservation identities of section \ref{charge-conservation-identities},
we obtain exactly the same expression as equation (\ref{e:bosonic-charge-identity}), except the
summation starts from $j=0$.  Thus, the existence of the spin $4$ current implies the existence of
a spin $0$ current with $\vev{\underline{22}_b0} \neq 0$. The action of charge $Q_4$ on $j_0$
is \begin{equation}[Q_4,j_0]=\partial^3 j_0+\partial j_2+\mbox{no overlap with
}\underline{22}_b\end{equation} We can consider charge conservation identities from the action of
$Q_4$ on $\vev{\underline{22}_f0}$ and $\vev{\underline{22}_v0}$.  However, $\vev{\underline{22}_f0}
= 0 = \vev{\underline{22}_v0}$. Thus, $\vev{\underline{22}_f 2} = 0 = \vev{\underline{22}_v2}$.

So now, we may assume that $\vev{\underline{22}_b2} = 0$, and it suffices to show that if
$\vev{\underline{22}_v2} \neq 0$, then $\vev{\underline{22}_f2} = 0$.  In this case, by hypothesis,
the quasi-bilocal $V_{--}$ is nonzero. The results of the previous section tell us that the three
point function of the vector quasi-bilocal is proportional to:
\begin{equation}
\vev{V_{--}(x_1,x_2)V_{--}(x_3,x_4)V_{--}(x_5,x_6)} \propto\frac{\left(x_{13}^+\right)^2 \left(
x_{25}^+\right)^2\left( x_{46}^+\right)^2}{d_{13}^6d_{25}^6d_{46}^6}+perm.\label{e:vecstruc}
\end{equation}
and this precisely coincides with the three-point function of the free field operator $v_{--}(x_1,x_2) =
:F_{-\alpha}(x_1)F_{-\alpha}(x_2):$
\begin{equation}
\vev{V_{--}(x_1,x_2)V_{--}(x_3,x_4)V_{--}(x_5,x_6)} \propto \vev{v_{--}(x_1,x_2)v_{--}(x_3,x_4)v_{--}(x_5,x_6)}
\label{e:vvv}
\end{equation}
Now, take $x_1$ and $x_2$ very close together and expand both sides of this equation
in powers of $(x_1-x_2)$.  The zeroth order term of $v$ is clearly the normal ordered product
$:F_{-\alpha}(\frac{x_1+x_2}{2})F_{-\alpha}(\frac{x_1+x_2}{2}):$ - this is precisely the free field
stress tensor.  On the other hand, consider the vector analogue of (\ref{e:decomposition})
\begin{align}
V_{--}(x_1,x_2) &= \sum_{\text{even } s\ge 2} v^s_{--}(x_1,x_2) \\
v^s_{--}(x_1,x_2) &= \sum_{k,l} c_{kl}(x_1-x_2)^k\pd^lj_s\left(\frac{x_1+x_2}{2}\right)
\end{align}
We know that $V_{--}$ transforms like a product of free fields, so comparing the conformal
dimension of the left and right hand side yields the constraing that $s+l-k = 2$.  Since we want to
extract the $k=0$ piece, this forces $l = 0$ and $s = 2$.  Repeating the same procedure for the
pairs of coordinates $(x_3,x_4)$ and $(x_5,x_6)$, we obtain the desired result:
\begin{align} 
\vev{222} &= \vev{222}_{\text{free vector}} \\
\Rightarrow \vev{\underline{22}_f2}&=\vev{\underline{22}_b2} = 0
\end{align}
as required. Therefore, since the stress-energy tensor is unique,
\begin{align}
\vev{222}_b&\neq0 \Rightarrow &\vev{222}_f&=0,& \vev{222}_v&=0,& \underline{j_2j_2}_b=&\sum_{k=0}^{\infty} \left[j_{2k}\right],& \underline{j_2j_2}_f&=0,&\underline{j_2j_2}_v &=0,
\\
\vev{222}_f&\neq0 \Rightarrow &\vev{222}_b&=0, &\vev{222}_v&=0,& \underline{j_2j_2}_f=&\sum_{k=1}^{\infty} \left[j_{2k}\right],& \underline{j_2j_2}_b&=0,&\underline{j_2j_2}_v &=0,
\\
\vev{222}_v&\neq0 \Rightarrow &\vev{222}_b&=0,& \vev{222}_f&=0,& \underline{j_2j_2}_v=&\sum_{k=1}^{\infty} \left[j_{2k}\right],& \underline{j_2j_2}_b&=0,&\underline{j_2j_2}_f &=0,
\end{align}
where square brackets denotes currents and their descendants.  This establishes the claim that the
three-point function of the stress tensor coincides with the answer for some free theory.

To obtain all the other correlation functions, we may expand equation (\ref{e:vvv}) to higher
orders in $x_1-x_2$, and use the correlation functions obtained at lower orders to fix the ones that
appear at higher orders.  For example, at second order in $x_1-x_2$, $v_{--}$ is
$(x_1-x_2)^2 \left(:\pd^2 F_{-\alpha}\left(\frac{x_1+x_2}{2})F_{-\alpha}(\frac{x_1+x_2}{2}\right):+:\pd F_{-\alpha}\left(\frac{x_1+x_2}{2})\pd F_{-\alpha}(\frac{x_1+x_2}{2}\right)\right)$, 
and $V_{--}$ contains terms involving only the spin $2$, $3$, and $4$ currents.  
Using our answers for $\vev{222}$ and our knowledge that $\vev{223} = 0$, we can then
fix $\vev{224}$ to agree with the free field theory.  This procedure recursively fixes all the 
correlators in the free vector sector.  The argument flows identically for the free bosonic and free
fermionic sectors, except that the zeroth order term will not fix $\vev{222}$, but some lower-order
current.  For example, in the bosonic theory, the zeroth order term will fix $\vev{000}$, and one
will need to carry out the power series expansion to higher orders in order to fix the correlators
of the higher-spin conserved currents.

Then, one could consider correlation functions that have indices set to values other than minus.
This works in exactly the same way, since the operator product expansion of two currents with minus
indices will contain currents with other indices.  This has the effect of doubling the number of
bilocals required to build a correlation function, since we need to take an extra OPE to fix the index
structure.  Thus, an $n$-point function with non-minus indices can be fixed from $2n$ bilocals.
Thus, we have fixed every correlation function from currents at appear in successive OPE's of two stress
tensors, including those of every higher-spin current.

The last thing to prove is that the normalization of the correlation functions matches the
normalization for some free theory.  For example, in the theory of $N$ free bosons, the 
two-point function of $\sum_{i=1}^N :\phi_i\phi^*_i:$ will have overall coefficient $N$.  
The same is true for the fermionic and vector cases.  One might wonder if the overall coefficient
$\tilde{N}$ of the quasi-bilocal could be non-integer, which would imply that it could not coincide
with any theory of $N$ free bosons.  We will now show that this is not possible.  We start with the
bosonic case, which works exactly as in \cite{Maldacena:2011jn} without modification:

In a theory of $N$ free bosons, consider the operator
\[\mathcal{O}_q = \delta^{[i_1, \dots,i_q]}_{[j_1,\dots,j_q]}
(\phi^{i_1}\pd\phi^{i_2}\dots\pd^{q-1}\phi^{i_q})
(\phi^{j_1}\pd\phi^{j_2}\dots\pd^{q-1}\phi^{j_q})\]
Here, $\delta$ is the totally antisymmetric delta function that arises from a partial contraction of
$\epsilon$ symbols:
\[\delta^{[i_1,\dots,i_q]}_{[j_1,\dots,j_q]} \propto 
\epsilon^{i_1\dots,i_q, i_{q+1}\dots i_N}\epsilon_{j_1\dots,j_q, i_{q+1}\dots i_N} \]
Upon contracting indices, this delta function ensures that $\mathcal{O}_q$ is a sum of products of
$q$ bilinear operators. Now, consider the norm of the state that $\mathcal{O}_q$ generates.  This
is computed by the two point function $\vev{\mathcal{O}_q\mathcal{O}_q}$.  The contractions of the
bilinears means that this correlator is a polynomial in $N$ of order $q$, and it is
not hard to see from explicit calculation that the correlation function vanishes at $q \ge N$.  So
we know all the roots of the polynomial, and hence the correlation function is proportional to
$N(N-1)\dots(N-(q-1))$.  Now, consider an analytic continuation of this correlator to 
non-integer $\tilde{N}$.  By taking $q = \lfloor N \rfloor + 2$, we find that this product is
negative, which is impossible for the norm of a state.  Since the correlators of $\mathcal{O}_q$ are
forced to agree with the correlators of some operator in the full CFT, we conclude that the
normalization $\tilde{N}$ of the scalar quasi-bilocals must be an integer.

The same argument can be ran in the vector case for an operator defined similarly:
\[\mathcal{O}_q = \delta^{[i_1, \dots,i_q]}_{[j_1,\dots,j_q]}
(c^{i_1}\pd c^{i_2}\dots \pd^{q-1}c^{i_q})
(\bar{c}^{j_1}\pd\bar{c}^{j_2}\dots\pd^{q-1}\bar{c}^{j_q})\]
where we have suppressed the spinor indices of $c$ and $\bar{c}$, the self-dual and anti-self
dual parts of $F$ defined before equation (\ref{e:c-definition}).  We again conclude that the
normalization constant $\tilde{N}$ must be an integer.

The construction in the fermionic case is somewhat simpler.  We know $j_2$ appears in $F_-$, and we
can define an operator $\mathcal{O}_q = (j_2)^q$ by extracting the term in the operator product
expansion of $q$ copies of $j_2$ whose correlation functions coincide with the free fermion operator
$(j_2)^q_{\text{free}}$.  In the theory of $N$ free fermions, $j_2 = \sum_i
(\pd\psi_i)\gamma_-\bar{\psi_i} - \psi_i\gamma_-(\pd\bar\psi_i)$, where here $i$ is the flavor index
for the $N$ fermions.  By antisymmetry of the fermions, we know that $\mathcal{O}_q$ will be zero if
$q \ge N$.  Then, as in the bosonic case, we can consider the norm of the state that $\mathcal{O}_q$
generates, which is computed by $\vev{\mathcal{O}_q\mathcal{O}_q}$, and the rest of the argument
runs as before.  Thus, the normalization $\tilde{N}$ of the fermionic bilocals must be an integer.

\section{Conclusions}
In this paper, we have shown that in a unitary conformal field theory in four dimensions with
a unique stress tensor and a symmetric conserved current of spin higher than $2$, the three-point
function of the stress tensor must coincide with the three-point function of the stress tensor in
either a theory of free bosons, free fermions, or free vector fields.  This then implies that all
the correlation functions of symmetric currents of the theory coincide with the those in the
corresponding free field theory.  Our technique was to use a set of appropriate lightcone limits and
the spinor-helicity formalism to transform the data of certain key Ward identities into simple
polynomial equations.  Even though we could not directly solve for the coefficients in these
identities like in three dimensions, we were nevertheless able to show that the only solution these
Ward identities admit is the one furnished by the appropriate free field theory.  This was the key
step that allowed us to defined bilocal operators which were used to show that the three-point
function of the stress tensor must agree with a free field theory.  This in turn fixed all the other
correlators of the theory to agree with those in the same free field theory.  These results can be
understood as an extension of the techniques and conclusions of \cite{Maldacena:2011jn} from three
dimensions to four dimensions.

We stress that our classification into the bosonic, fermionic, and vector free field theories
depends sharply on our assumption that a unique stress tensor exists.  Other free field theories
with higher spin symmetry exist in four dimensions, such as a theory of free gravitons.  This
theory, however, does not have a stress tensor, and we make no statement about how the correlation
functions of such theories are constrained, and analogously for theories with many stress tensors.
Moreover, we have not computed correlation functions or commutators for asymmetric currents and
charges.  In \cite{Boulanger:2011se}, it was shown that if one considers the possible algebras of
charges in theories that contain asymmetric currents, a one-parameter family of algebras exists.  
This may suggest the existence of nontrivial higher-spin theories, though our result indicates
that at least the subalgebra generated by the symmetric currents must agree with free field theory.

Unfortunately, our techniques do not easily generalize to higher dimensions, where the
spinor-helicity formalism does not exist, and where representation theoretic difficulties pose a
greater challenge.  For this case, additional techniques will be necessary to establish the
relevant identities, assuming the same proof strategy still works.

\acknowledgments
We would like to thank J.~Maldacena and A.~Zhiboedov for countless helpful discussions and invaluable
guidance on this project. We would also like to thank E.~Skvortsov and D.~Ponomarev for their help
in understanding their recent paper.

The work of VA was supported by the National Science Foundation under Grant No. PHY-0756966.
The work of KD was supported by the National Science Foundation Graduate Research Fellowship under
Grant No.~DGE 1148900.

\appendix\label{appendix}
\section{Representation theory of the Lorentz group}\label{representation-theory}
\subsection{Spinor notation}
Here, we will review some basic facts that allow us to translate equations written in vector indices
to spinor indices.  The key fact is that the Euclidean Lie algebra $\mathfrak{so}_4$ is isomorphic
to $\mathfrak{su}_2 \times \mathfrak{su}_2$.  This isomorphism leads to the classification of the
representations of the Lorentz group in terms of pairs of half-integers $(p,q)$, which are the
eigenvalues of the $J_z$ rotation operator on each $\mathfrak{su}_2$ factor.
This isomorphism of algebras also induces a isomorphism of representations between the vector
representation $V$ (which is the $(1/2,1/2)$ representation)
and the product $S^- \times S^+$ between the left-handed and right-handed Weyl
representations (which transform as $(1/2,0)$ and $(0,1/2)$, respectively, and are expressed as
spinors $\lambda_a$, $a=1,2$, for $S^-$ and $\tilde\lambda_{\dot a}$, $\dot a= \dot 1,\dot 2$ for
$S^+$).  This isomorphism is given by the ordinary Pauli matrices:
\begin{align}
\sigma: V\rightarrow S^-\times S^+,\qquad \bar\sigma : S^-\times S^+\rightarrow V.\\
p_\mu \sigma^\mu_{a\dot a}=p_{a\dot a},\qquad p_{a\dot a}\left(\bar \sigma^\mu\right)^{\dot a a}=p^\mu.
\end{align}
For example, this map yields the correspondences $x_{1\dot 1} = x_-$, $x_{2 \dot 2} = x_+$, etc.

For higher-dimensional symmetric representations (for example, a completely symmetric tensor
$T_{\mu_1\dots\mu_n}$, which transforms in the $(n/2, n/2)$ representation), the map 
to spinor indices is the natural extension of the previous formula:
\begin{equation}
T_{\mu_1\dots\mu_n}\sigma^{\mu_1}_{a_1\dot a_1}\dots\sigma^{\mu_n}_{a_n \dot a_n} = \tilde
T_{a_1\dots a_n \dot a_1\dots \dot a_n}. 
\end{equation}
Tensors in the $(p,q)$ representationwith $p \neq q$ have the property that they are antisymmetric
instead of symmetric under interchange of some of their indices.  In spinor indices,
$T_{a_1\dots a_{2p} \dot a_1\dots \dot a_{2q}}$ will be also antisymmetric under the permutation 
of some $a_i$ and $a_j$, but the only object with two indices that is totally antisymmetric under 
the permutation of two indices is $\epsilon_{ab}$.  In the case where $p>q$, we will then obtain
\begin{equation}
T_{\mu_1\dots \mu_{2p}}\sigma^{\mu_1}_{a_1\dot a_1}\dots\sigma^{\mu_{2p}}_{a_{2p} \dot a_{2p}}
=T_{a_1\dots a_{2p}\dot a_1\dots \dot a_{2q}} \epsilon_{\dot a_{2q+1}\dot a_{2q+2}}\dots 
\epsilon_{\dot a_{2p-1}\dot a_{2p}}.
\end{equation}
Put another way, the symmetrizers for two tensor representations are the same
\begin{equation}
\Sigma\left (\prod V\right)=\Sigma'\left(\prod S^+,\prod S^-\right).
\end{equation}
Translating from spinor to vector indices follows the same rules.  For example, for a $(p,q)$ tensor
with $p > q$, we have
\begin{equation}
T_{a_1\dots a_{2p}\dot a_1\dots \dot a_{2q}} \epsilon_{\dot a_{2q+1}\dot a_{2q+2}}\dots 
\left(\bar\sigma^{\mu_1}\right)^{\dot a_1 a_1}\dots\left(\bar\sigma^{\mu_{2p}}\right)^{\dot a_{2p} a_{2p}}
= T^{\mu_1\dots\mu_{2p}}.
\end{equation}
\subsection{Spin}
Here, we will state how the eigenvalues $(p,q)$ of the $J_z$ operators of the $\mathfrak{su}_2$ 
factors are related to rotation operators in $\mathfrak{so}_4$.  Recall $\mathfrak{su}(2)$ has 
three generators $(J_x,J_y,J_z)$, while the $\mathfrak{so}(4)$ has 6 generators; three rotations 
$J_{ij}$ and three boosts $J_{0i}$.  In Minkowski signature, the correspondence between these 
two sets of generators is given by:
\begin{align}
J_k^{(1)}&=\epsilon_{ijk}J_{ij}+J_{0k},\\
J_k^{(2)}&=\epsilon_{ijk}J_{ij}-J_{0k}.
\end{align} 
We are interested in the $z$ component:
\begin{align}
J_z^{(1)}&=J_{12}+iJ_{03},\\
J_z^{(2)}&=J_{12}-iJ_{03}.
\end{align} 
This leads to
\begin{align}
J_{12}&=\frac{1}{2}\left(J_z^{(1)}+J_z^{(2)}\right),\\
iJ_{03}&=\frac{1}{2}\left(J_z^{(1)}-J_z^{(2)}\right).
\end{align} 
In Euclidean signature:
\begin{align}
J_{12}&=\frac{1}{2}\left(J_z^{(1)}+J_z^{(2)}\right),\\
J_{34}&=\frac{1}{2}\left(J_z^{(1)}-J_z^{(2)}\right).
\end{align} 

It means that we have two planes and rotations in one plane correspond to one spin while rotations
into another plane correspond to another plane.

\section{Form factors as Fourier transforms of correlation functions}\label{fourier-transforms}
Let's start with the bosonic case. Fourier transforming the lightcone limit of the three point
functions with respect to $x_1^-$ and $x_2^-$ and using the fact that in the lightcone limit we 
have $x_1^+ = x_2^+$ and $\vec{y}_1 = \vec{y}_2$, we obtain:
\begin{align}
&\pd_1^{s-i}\pd_2^i\vev{\phi(x_1)\phi^*(x_3)}\vev{\phi(x_3)\phi^*(x_2)} 
\\
&\longrightarrow i^s(p_1^+)^{s-i}(p_2^+)^i \int dx_1^-dx_2^- e^{ip_1^+x_1^-}e^{ip_2^+x_2^-}
\frac{1}{x_{13}^+x_{13}^- + \vec{y}_{13}^2} \frac{1}{x_{23}^+x_{23}^- + \vec{y}_{23}^2}
\\
&= i^s(p_1^+)^{s-i}(p_2^+)^i \int dx_1^-dx_2^- e^{ip_1^+x_1^-}e^{ip_2^+x_2^-}
\frac{1}{x_{13}^+\left(x_{13}^- + \frac{\vec{y}_{13}^2}{x_{13}^+} - i\epsilon \right)} 
\frac{1}{x_{13}^+\left(x_{23}^- + \frac{\vec{y}_{13}^2}{x_{13}^+} - i\epsilon \right)}\\
&=-4\pi^2i^s \frac{(p_1^+)^{s-i}(p_2^+)^i}{(x_{13}^+)^2}e^{ip_1^+\bar{x} + ip_2^+\bar{x}}
\\
&\propto (p_1^+)^{s-i}(p_2^+)^i \times \text{(a nonsingular function)}
\end{align}
Here, $\bar{x} = x_3^- - \frac{\vec{y}_{13}^2}{x_{13}^+}$ is the location of the poles in the integral.
Now, since $p \sim \pi\tilde{\pi}$, we find that $\vev{\phi\phi^*s_3}$ scales like 
$\sum c_ip_1^{s-i}p_2^i \sim \sum c_i \pi_1^{s-1}\tilde{\pi}_1^{s-i}\pi_2^i\tilde{\pi}_2^i$.
Since spinors have to contract with other spinors of the same helicity, and the only other spinors
to contract with are $\lambda$ and $\tilde{\lambda}$, $\vev{\phi\phi^*s_3}$ scales like $\sum
c_i(x\tilde{x})^{s-i}(y\tilde{y})^i$. Hence, as claimed, $F_s$ transforms into a polynomial 
in $x\tilde{x}$ and $y\tilde{y}$ multiplied by some universal nonsingular function which can be
divided out in the charge conservation identities. That is, the Fourier transform does not render
some of the charge conservation identities trivial, as we might have feared.

For the fermion, the Fourier transform required involves a non-meromorphic square root:
\begin{align}
&\pd_1^{s-i-1}\pd_2^i\vev{\psi(x_1)\bar{\psi}(x_3)}\vev{\psi(x_3)\bar{\psi}(x_2)} 
\\
&\longrightarrow i^s(p_1^+)^{s-i-1}(p_2^+)^i \int dx_1^-dx_2^- e^{ip_1^+x_1^-}e^{ip_2^+x_2^-}
\frac{1}{(x_{13}^+x_{13}^- + \vec{y}_{13}^2)^{3/2}} \frac{1}{(x_{23}^+x_{23}^- +
\vec{y}_{23}^2)^{3/2}}
\\
&= i^s\frac{(p_1^+)^{s-i-1}(p_2^+)^i}{(x_{13}^+)^3} \int dx_1^-dx_2^- e^{ip_1^+x_1^-}e^{ip_2^+x_2^-}
\frac{1}{\left(x_{13}^-+\frac{\vec{y}_{13}^2}{x_{13}^+}\right)^{3/2}} 
\frac{1}{\left(x_{23}^-+\frac{\vec{y}_{13}^2}{x_{13}^+}\right)^{3/2}}
\end{align}
To deal with the branch points, we analytically continue both integrals. Both have the schematic
form $\int_{-\infty}^\infty \frac{e^{ipx}}{x^{3/2}}$. Depending on the sign of $p$, we move the
branch cut up or down by $i\epsilon$ so that $ipx < 0$ at the branch point and then take $s = -ix$.
Then the integrals become a simple product of $\Gamma$ functions, which are nonsingular, and so $F$
has the desired property.

For the vector case, we have double poles which, as in the boson case, are nonsingular.
\begin{align}
&\pd_1^{s-i-2}\pd_2^i\vev{c(x_1)\bar{c}(x_3)}\vev{c(x_3)\bar{c}(x_2)} 
\\
&\longrightarrow i^s(p_1^+)^{s-i-2}(p_2^+)^i \int dx_1^-dx_2^- e^{ip_1^+x_1^-}e^{ip_2^+x_2^-}
\frac{1}{(x_{13}^+x_{13}^- + \vec{y}_{13}^2)^2} \frac{1}{(x_{23}^+x_{23}^- +
\vec{y}_{23}^2)^2}
\\
&=-4\pi^2i^{s+2} \frac{(p_1^+)^{s-i-1}(p_2^+)^{i+1}}{(x_{13}^+)^2}e^{ip_1^+\bar{x} + ip_2^+\bar{x}}
\\
&\propto (p_1^+)^{s-i-2}(p_2^+)^{i} \times p_1^+p_2^+ \times \text{(a nonsingular function)}
\end{align}
Again, $F$ is nonsingular, as required.

\section{Conformal invariants}\label{invariants}
Conformal invariance is Poincare invariance with invariance under the action of the special
conformal transformation and dilatation operators. Therefore, in order to build conformally invariant
objects, we should write down structures which are manifestly Poincare invariant structure and are 
homogeneous under rescaling. 

In analogy with \cite{Giombi:2011rz} and \cite{Todorov:2012xx}, we can easily identify the following
conformal invariants: \begin{align}
P_{i,j}&=\lambda_i \check {\bf x}_{ij} \tilde\lambda_j,
\\
Q_i&= -\lambda_i \left( \check{\bf x}_{i,i+1}-\check{\bf x}_{i,i+2} \right) \tilde \lambda_i,
\end{align}
where
\begin{equation}
\check {\bf x}=\frac{x^\mu \left(\sigma_\mu\right)_{a\dot a}}{x^2}, \qquad \sigma^\mu=\left(
\mathds {1},\vec\sigma,\right),\,\bar \sigma_\mu = \left( \mathds {1},\vec \sigma\right).
\end{equation}
We have 9 these invariants. Let's make sure that they form a complete basis. For each point we have
4 degrees of freedom that correspond to the coordinates ${\bf x}_i$, and $2+2$ degrees of freedom that
correspond to the spinor variables $\lambda_i,\tilde\lambda_i$. The dimension of the conformal group is
15. Therefore, in total, for the three-point function we expect have \begin{equation}3(4+2+2)-15=9\end{equation}
degrees of freedom. This number coincides with the number of $P_{i,j},Q_i$. One could wonder why did
not take into account the gauge symmetry for spinors \begin{equation}
\lambda_i \longrightarrow z_i \lambda_i,\quad \tilde\lambda_i\longrightarrow \frac{1}{z_i}
\tilde\lambda_i, \quad z_i\in \mathds{C}^*.\end{equation}
We did not take this into account because our conformal invariants are not invariant under this 
transformation. Indeed, 
\begin{equation}P_{ij}\longrightarrow \frac{z_i}{z_j}P_{ij}.\end{equation}
If we would like to take into account these symmetries we should consider a different set of invariants
\begin{equation}Q_i=-V_i,\quad H_{ij}=2P_{ij}P_{ji}.\end{equation}
These invariants are also invariants under the gauge transformation and it means that the count of
degrees of freedom is different now 
\begin{equation}3(4+2+2-1)-15=6.\end{equation}
\section{Uniqueness of three-point functions in the vector lightcone limit}\label{vector-uniqueness}
Our goal in this section is to show that the free vector solution for the lightcone limit of
three-point functions explained in section \ref{form-factors} is indeed unique, at least in the
lightcone limit.

Note that Lorentz symmetry constrains the propagator of spin $j$ field to be of the form
\begin{equation}
\vev{\psi_{-j}(x)\bar\psi_{-j}(0)}\propto (x^+)^{2j}.
\end{equation}
Generically, according to \cite{Zhiboedov:2012bm}, the most generic conformally invariant expression
one can write down for a three-point function of symmetric conserved currents with vector-type 
coordinate dependence is:
\begin{multline}
\vev{j_{s_1}j_{s_2}j_{s_3}}=\frac{1}{x_{12}^2 x_{23}^2x_{13}^2} \times
\\
\times\sum\limits_{a,b,c}(\Lambda_1^2\alpha_{a,b,c}+\Lambda_2\beta_{a,b,c})\left(P_{12} P_{21}\right){}^a Q_1^b \left(P_{23} P_{32}\right){}^c \left(P_{13} P_{31}\right){}^{-a-b+s_1} Q_2^{-a-c+s_2} Q_3^{a+b-c-s_1+s_3}
\end{multline}
where the $\alpha_{a,b,c}$ and $\beta_{a,b,c}$ are free coefficients, and the $\Lambda_i$ are defined as: 
\begin{align}
\Lambda_1&= Q_1Q_2Q_3+\left[ Q_1 P_{23}P_{32}+Q_2P_{13}P_{31}+Q_3P_{12}P_{21}\right],
\\
\Lambda_2&=8P_{12}P_{21}P_{23}P_{32}P_{13}P_{31}.
\end{align}
Here, the $P$ and $Q$ invariants are defined as in \cite{Giombi:2011rz} and \cite{Todorov:2012xx},
and the basic properties of these invariants are compiled in appendix \ref{invariants}. However, for
the choice of polarization vector $\epsilon^\mu=\epsilon^-$ there is a nontrivial relation:
\begin{equation}
\Lambda_2\big|_{\epsilon^\mu_i=\epsilon^-}=-2\Lambda_1^2\big|_{\epsilon^\mu_i=\epsilon^-}, \quad \Lambda_1\big|_{\epsilon_i^\mu=\epsilon^-}=\frac{1}{4}\frac{x^+_{12}x^+_{23}x^+_{13}}{x^2_{12}x_{23}^2x_{13}^2} (\epsilon^-)^3.
\end{equation}
Therefore, in the case $\epsilon^\mu=\epsilon^-$ the expression for this three-point function
greatly simplifies. Instead of having two sets of undetermined coefficients $c_a$ and $d_a$, one
can combine the $\Lambda_i$'s into a single prefactor $\alpha_1 \Lambda_1^2+\alpha_2\Lambda_2$,
where the $\alpha_i$ are arbitrary and can be chosen to be convenient; to produce exact agreement
with the canonically normalized free-vector theory, we will choose $\alpha_1 = 1$ and $\alpha_2 =
\frac{1}{2(d-2)} = \frac{1}{4}$. Now, we take the lightcone limit, which corresponds to the point
where
\begin{equation}
P_{23}P_{32}=0,\quad Q_1=-\left( \frac{P_{13}P_{31}}{Q_3}+\frac{P_{12}P_{21}}{Q_2}\right)
\end{equation}
in $P_{ij}, Q_i$ space. Then, the three-point function reduces to
\begin{equation}
\vev{j_{s_1}\underline{j_{s_2}j_{s_3}}_v}=\frac{\Lambda_1^2+\Lambda_2/(2(d-2))}{x_{12}^2 x_{23}^2x_{13}^2} \sum\limits_{a=0}^{s_1-2}c_a\left( P_{12}P_{21}\right)^a \left( P_{13}P_{31}\right)^{s_1-2-a} Q_{2}^{s_2-a}Q_{3}^{S_3-s_1+a}, 
\end{equation}
Now, the $c_a$ can be fixed demanding that all currents are conserved. The result is given by the
following recurrence relation, with $c_0 = 1$:
\[ \frac{c(a+1)}{c(a)} = \frac{(s_1-2-a)(s_1 + \frac{d-4}{2}-a)(s_2+a+\frac{d-2}{2})}
{(a+1)(a+\frac{d-2}{2}+2)(s_1+s_3+\frac{d-4}{2}-2-a)} \]
This solution exactly coincides with the free vector solution, as required.
\section{Uniqueness of $\vev{s22}$ for $s \ge 4$}\label{uniqueness-of-22s}
Let us use the following definition of double brackets
\begin{equation}
\vev{j_{s_1}j_{s_2}j_{s_3}}=\frac{\vev{\vev{j_{s_1}j_{s_2}j_{s_3}}}}{{x_{12}}^{d-2}{x_{23}}^{d-2}{x_{13}}^{d-2}}.
\end{equation}
 Using conformal invariants we can write the most general expression for a conformal invariant correlation function
\begin{multline}
\vev{\vev{j_sj_2j_2}} = V_1^{s-4} \Big[ a_1 H_{1,2}^2 H_{1,3}^2+ a_2 \left(V_1 V_2 H_{1,2} H_{1,3}^2+V_1
V_3 H_{1,2}^2 H_{1,3}\right) + a_3 V_1^2 H_{1,2} H_{1,3} H_{2,3}+
\\
+a_4 \left( V_1^2 V_3^2 H_{1,2}^2+V_1^2V_2^2 H_{1,3}^2 \right) +a_5 V_1^2V_2 V_3 H_{1,2} H_{1,3}+
\\
+a_6 \left( V_1^3V_2 H_{1,3} H_{2,3}+ V_1^3V_3 H_{1,2} H_{2,3}\right)+a_7 \left( V_1^3V_2 V_3^2
H_{1,2}+V_1^3V_2^2 V_3 H_{1,3}\right)+
\\
a_8 V_1^4 H_{2,3}^2+a_9V_1^4 V_2 V_3 H_{2,3}+a_{10} V_1^4V_2^2 V_3^2 \Big].
\end{multline}
From conservation condition in $d=4$ it follows that 
\begin{align}
a_1&= -\frac{a_7 (s-3) (s-1) (s-2)^2}{32 (s+1) (s+4)}+\frac{a_4 (s-5) (s-3) s (s-2)}{8 (s+1) (s+4)}+\frac{a_5 (s-3) (s-2)}{8 (s+4)},
\\
a_2&= -\frac{a_4 (s-2)^2}{s+4}+\frac{a_7 (s-1) (s-2)}{4 (s+4)}-\frac{a_5 (s-2)}{2 (s+4)},
\\
a_3&= -\frac{8 a_4 \left(s^2-3 s-1\right)}{(s+1) (s+4)}+\frac{a_5 (s-8)}{2 (s+4)}+\frac{a_7 (s-1) (2 s-1)}{(s+1) (s+4)},
\\
a_6&= \frac{12 a_4 (s-2)}{(s-1) (s+4)}+\frac{6 a_5}{(s-1) (s+4)}+\frac{a_7 (s-2)}{2 (s+4)},
\\
a_8&= \frac{a_7 (s-2) \left(s^2+11 s-2\right)}{4 s (s+1) (s+4)}-\frac{6 a_4 (s-5)}{(s+1) (s+4)}+\frac{a_5 (s-2)}{s (s+4)},
\\
a_9&= \frac{a_7 \left(s^2+8 s-8\right)}{s (s+4)}-\frac{24 a_4 (s-2)}{(s-1) (s+4)}+\frac{4 a_5 (s-2) (s+2)}{(s-1) s (s+4)},
\\
a_{10}&= \frac{a_7 \left(s^2+8 s+4\right)}{s (s+4)}-\frac{24 a_4 (s+1)}{(s-1) (s+4)}+\frac{4 a_5 (s+1) (s+2)}{(s-1) s (s+4)}.
\end{align}
Therefore, $\vev{\vev{j_sj_2j_2}}_v$ depends only on three parameters. The bosonic light-cone limit of this function is zero if 
\begin{equation}a_5= \frac{a_7 (s-2) (s-1)}{4 (s+1)}-\frac{a_4 (s-5) s}{s+1}.\end{equation}
While the fermionic light-cone limit of this function is also zero if 
\begin{equation}a_4= \frac{a_7}{4}.\end{equation}
Therefore $\vev{\vev{s22}}_v$ depends only on one parameter or in other words it is unique up to a rescaling\footnote{In  \cite{Stanev:2012nq} it was proven that there are only three structures for $\vev{\vev{22s}}$ in d=4.} 
\begin{equation}
\vev{\vev{j_sj_2j_2}}_v \propto V_1^{s-2}\Big[H_{12}^2 V_3^2+\left(H_{23} V_1+V_2 \left(H_{13}+2 V_1 V_3\right)\right){}^2+H_{12} \left(H_{13}+2 V_1 V_3\right) \left(H_{23}+2 V_2 V_3\right)\Big], 
\end{equation}
or for arbitrary $d$
\begin{align}
\vev{\vev{j_sj_2j_2}}_v &= V_1^{s-2} \left[\left(H_{23} V_1+H_{13} V_2+H_{12} V_3+2 V_2 V_3
V_1\right){}^2+\frac{2}{(d-2)} H_{12}H_{13} H_{23}\right]\nonumber\\
&=V_1^{s-2}\left[ \Lambda_1^2+\frac{1}{2(d-2)}\Lambda_2\right].
\end{align}
This formula coincides with the expression that was proposed in \cite{Zhiboedov:2012bm}, however now it has been proven that this structure is unique.

\bibliographystyle{JHEP}
\bibliography{final-writeup}

\providecommand{\href}[2]{#2}\begingroup\raggedright\begin{thebibliography}{10}

\bibitem{Maldacena:2011jn}
J.~Maldacena and A.~Zhiboedov, {\it {Constraining Conformal Field Theories with
  A Higher Spin Symmetry}},  {\em J.Phys.} {\bf A46} (2013) 214011,
  [\href{http://xxx.lanl.gov/abs/1112.1016}{{\tt arXiv:1112.1016}}].

\bibitem{Konstein:2000bi}
S.~Konstein, M.~Vasiliev, and V.~Zaikin, {\it {Conformal higher spin currents
  in any dimension and AdS / CFT correspondence}},  {\em JHEP} {\bf 0012}
  (2000) 018, [\href{http://xxx.lanl.gov/abs/hep-th/0010239}{{\tt
  hep-th/0010239}}].

\bibitem{Vasiliev:2003ev}
M.~Vasiliev, {\it {Nonlinear equations for symmetric massless higher spin
  fields in (A)dS(d)}},  {\em Phys.Lett.} {\bf B567} (2003) 139--151,
  [\href{http://xxx.lanl.gov/abs/hep-th/0304049}{{\tt hep-th/0304049}}].

\bibitem{Vasiliev:2004qz}
M.~Vasiliev, {\it {Higher spin gauge theories in various dimensions}},  {\em
  Fortsch.Phys.} {\bf 52} (2004) 702--717,
  [\href{http://xxx.lanl.gov/abs/hep-th/0401177}{{\tt hep-th/0401177}}].

\bibitem{Maldacena:1997re}
J.~M. Maldacena, {\it {The Large N limit of superconformal field theories and
  supergravity}},  {\em Adv.Theor.Math.Phys.} {\bf 2} (1998) 231--252,
  [\href{http://xxx.lanl.gov/abs/hep-th/9711200}{{\tt hep-th/9711200}}].

\bibitem{Gubser:1998bc}
S.~Gubser, I.~R. Klebanov, and A.~M. Polyakov, {\it {Gauge theory correlators
  from noncritical string theory}},  {\em Phys.Lett.} {\bf B428} (1998)
  105--114, [\href{http://xxx.lanl.gov/abs/hep-th/9802109}{{\tt
  hep-th/9802109}}].

\bibitem{Witten:1998qj}
E.~Witten, {\it {Anti-de Sitter space and holography}},  {\em
  Adv.Theor.Math.Phys.} {\bf 2} (1998) 253--291,
  [\href{http://xxx.lanl.gov/abs/hep-th/9802150}{{\tt hep-th/9802150}}].

\bibitem{Klebanov:2002ja}
I.~Klebanov and A.~Polyakov, {\it {AdS dual of the critical O(N) vector
  model}},  {\em Phys.Lett.} {\bf B550} (2002) 213--219,
  [\href{http://xxx.lanl.gov/abs/hep-th/0210114}{{\tt hep-th/0210114}}].

\bibitem{Sezgin:2003pt}
E.~Sezgin and P.~Sundell, {\it {Holography in {4D} (super) higher spin theories
  and a test via cubic scalar couplings}},  {\em JHEP} {\bf 0507} (2005) 044,
  [\href{http://xxx.lanl.gov/abs/hep-th/0305040}{{\tt hep-th/0305040}}].

\bibitem{Coleman:1967ad}
S.~R. Coleman and J.~Mandula, {\it {All possible symmetries of the S-matrix}},
  {\em Phys.Rev.} {\bf 159} (1967) 1251--1256.

\bibitem{Haag:1974qh}
R.~Haag, J.~T. Lopuszanski, and M.~Sohnius, {\it {All Possible Generators of
  Supersymmetries of the s Matrix}},  {\em Nucl.Phys.} {\bf B88} (1975) 257.

\bibitem{Boulanger:2013zza}
N.~Boulanger, D.~Ponomarev, E.~Skvortsov, and M.~Taronna, {\it {On the
  uniqueness of higher-spin symmetries in AdS and CFT}},
  \href{http://xxx.lanl.gov/abs/1305.5180}{{\tt arXiv:1305.5180}}.

\bibitem{Boulanger:2011se}
N.~Boulanger and E.~Skvortsov, {\it {Higher-spin algebras and cubic
  interactions for simple mixed-symmetry fields in AdS spacetime}},  {\em JHEP}
  {\bf 1109} (2011) 063, [\href{http://xxx.lanl.gov/abs/1107.5028}{{\tt
  arXiv:1107.5028}}].

\bibitem{Komargodski:2012ek}
Z.~Komargodski and A.~Zhiboedov, {\it {Convexity and Liberation at Large
  Spin}},  \href{http://xxx.lanl.gov/abs/1212.4103}{{\tt arXiv:1212.4103}}.

\bibitem{Stanev:1988ft}
Y.~Stanev, {\it {Stress-Energy tensor {and} U(1) Current Operator Product
  Expansions in Conformal QFT}},  {\em Bulg.J.Phys.} {\bf 15} (1988) 93--107.

\bibitem{Osborn:1993cr}
H.~Osborn and A.~Petkou, {\it {Implications of conformal invariance in field
  theories for general dimensions}},  {\em Annals Phys.} {\bf 231} (1994)
  311--362, [\href{http://xxx.lanl.gov/abs/hep-th/9307010}{{\tt
  hep-th/9307010}}].

\bibitem{Stanev:2012nq}
Y.~S. Stanev, {\it {Correlation Functions of Conserved Currents in Four
  Dimensional Conformal Field Theory}},  {\em Nucl.Phys.} {\bf B865} (2012)
  200--215, [\href{http://xxx.lanl.gov/abs/1206.5639}{{\tt arXiv:1206.5639}}].

\bibitem{Zhiboedov:2012bm}
A.~Zhiboedov, {\it {A note on three-point functions of conserved currents}},
  \href{http://xxx.lanl.gov/abs/1206.6370}{{\tt arXiv:1206.6370}}.

\bibitem{Stanev:2013qra}
Y.~S. Stanev, {\it {Constraining conformal field theory with higher spin
  symmetry in four dimensions}},  \href{http://xxx.lanl.gov/abs/1307.5209}{{\tt
  arXiv:1307.5209}}.

\bibitem{Nikolov:2000pm}
N.~M. Nikolov and I.~T. Todorov, {\it {Rationality of conformally invariant
  local correlation functions on compactified Minkowski space}},  {\em
  Commun.Math.Phys.} {\bf 218} (2001) 417--436,
  [\href{http://xxx.lanl.gov/abs/hep-th/0009004}{{\tt hep-th/0009004}}].

\bibitem{Mikhailov:2002bp}
A.~Mikhailov, {\it {Notes on higher spin symmetries}},
  \href{http://xxx.lanl.gov/abs/hep-th/0201019}{{\tt hep-th/0201019}}.

\bibitem{Cachazo:2005ga}
F.~Cachazo and P.~Svrcek, {\it {Lectures on twistor strings and perturbative
  Yang-Mills theory}},  {\em PoS} {\bf RTN2005} (2005) 004,
  [\href{http://xxx.lanl.gov/abs/hep-th/0504194}{{\tt hep-th/0504194}}].

\bibitem{Witten:2003nn}
E.~Witten, {\it {Perturbative gauge theory as a string theory in twistor
  space}},  {\em Commun.Math.Phys.} {\bf 252} (2004) 189--258,
  [\href{http://xxx.lanl.gov/abs/hep-th/0312171}{{\tt hep-th/0312171}}].

\bibitem{Anselmi:1999bb}
D.~Anselmi, {\it {Higher spin current multiplets in operator product
  expansions}},  {\em Class.Quant.Grav.} {\bf 17} (2000) 1383--1400,
  [\href{http://xxx.lanl.gov/abs/hep-th/9906167}{{\tt hep-th/9906167}}].

\bibitem{Dolan:2000ut}
F.~Dolan and H.~Osborn, {\it {Conformal four point functions and the operator
  product expansion}},  {\em Nucl.Phys.} {\bf B599} (2001) 459--496,
  [\href{http://xxx.lanl.gov/abs/hep-th/0011040}{{\tt hep-th/0011040}}].

\bibitem{Giombi:2011rz}
S.~Giombi, S.~Prakash, and X.~Yin, {\it {A Note on CFT Correlators in Three
  Dimensions}},  {\em JHEP} {\bf 1307} (2013) 105,
  [\href{http://xxx.lanl.gov/abs/1104.4317}{{\tt arXiv:1104.4317}}].

\bibitem{Todorov:2012xx}
I.~Todorov, {\it {Conformal field theories with infinitely many conservation
  laws}},  {\em J.Math.Phys.} {\bf 54} (2013) 022303,
  [\href{http://xxx.lanl.gov/abs/1207.3661}{{\tt arXiv:1207.3661}}].

\end{thebibliography}\endgroup
\end{document}